\documentclass[aps,prb,twocolumn,superscriptaddress,showpacs,amsmath,amssymb,longbibliography]{revtex4-1}

\usepackage{amsmath}
\usepackage{amsfonts}
\usepackage{amssymb}
\usepackage{graphicx}
\usepackage{color}

\usepackage{bm}
\usepackage{braket}
\usepackage{gensymb}
\usepackage{wasysym}

\usepackage{hyperref}

\newcommand{\calA}{\mathcal{A}}
\newcommand{\calU}{\mathcal{U}}
\newcommand{\calG}{\mathcal{G}}
\newcommand{\calS}{\mathcal{S}}
\newcommand{\trans}{\mathcal{T}}

\begin{document}

\title{$\mathrm{SU}(4)$-Symmetric Quantum Spin-Orbital Liquids on Various Lattices}

\author{Masahiko G. Yamada}
\email[]{myamada@mp.es.osaka-u.ac.jp}
\affiliation{Department of Materials Engineering Science, Osaka University, Toyonaka 560-8531, Japan.}
\affiliation{Institute for Solid State Physics, University of Tokyo, Kashiwa 277-8581, Japan.}
\author{Masaki Oshikawa}
\affiliation{Institute for Solid State Physics, University of Tokyo, Kashiwa 277-8581, Japan.}
\author{George Jackeli}
\altaffiliation[]{Also at Andronikashvili Institute of Physics, 0177
Tbilisi, Georgia.}
\affiliation{Max Planck Institute for Solid State Research, Heisenbergstrasse 1, D-70569 Stuttgart, Germany.}
\affiliation{Institute for Functional Matter and Quantum Technologies, 
University of Stuttgart, Pfaffenwaldring 57, D-70569 Stuttgart, Germany.}
\date{\today}

\begin{abstract}
An emergent $\mathrm{SU}(4)$ symmetry discovered in the microscopic model for $d^1$ honeycomb materials
[M.~G.~Yamada, M.~Oshikawa, and G.~Jackeli, Phys. Rev. Lett. \textbf{121}, 097201 (2018).]
has enabled us to tailor exotic $\mathrm{SU}(4)$ models in real materials.
In the honeycomb structure, the emergent $\mathrm{SU}(4)$ Heisenberg model
would potentially have a quantum spin-orbital liquid ground state due to the
\textit{multicomponent frustration}, and
we can expect similar spin-orbital liquids also in three-dimensinal versions of the honeycomb
lattice.  In such quantum spin-orbital liquids, both the spin and orbital degrees
of freedom become fractionalized and entangled together
due to the strong frustrated interactions between them.
Similarly to spinons in pure quantum spin liquids,
quantum spin-orbital liquids can host not only spinon excitations,
but also fermionic \textit{orbitalon} excitations at low temperature.
\end{abstract}

\maketitle

\section{Introduction}

The material realization of an $\mathrm{SU}(N)$ symmetry with $N>2$ was a long-standing problem.
The potential of an emergent $\mathrm{SU}(4)$ symmetry in spin-orbital
$d^1$ honeycomb materials~\cite{Yamada2018,Natori2018su4} has stimulated
research on various $\mathrm{SU}(4)$ models from two to three dimensions~\cite{Yamada2018},
including a prediction of a spinon-orbitalon Fermi surface~\cite{Natori2018su4}
in the three-dimensional (3D) case.  In these $d^1$ materials with one electron
in a $d$-shell,
the low-energy effective spin-orbital model becomes the $\mathrm{SU}(4)$ Heisenberg model,
which had been previously very difficult to be realized even in cold atomic systems~\cite{Cazalilla2014}.
Various quantum spin-orbital liquids (QSOLs) are indeed expected in such $\mathrm{SU}(4)$
models.

The $\mathrm{SU}(4)$ Heisenberg models have attractive advantages from a viewpoint
of frustrated magnetism.  One of the most intriguing features is that another type of
frustration called \emph{multicomponent frustration} exists even in bipartite lattices~\cite{Corboz2012}.
Triangular geometric frustration is not a necessary
condition for $\mathrm{SU}(4)$ spin-orbital liquids and thus we are able to discuss
several bipartite lattices~\cite{Corboz2012,Natori2018su4}
in this paper as potential hosts of the spin-orbital liquids.
Additionally we will also discuss the
broken $\mathrm{SU}(4)$ symmetry on the triangular lattice and its consequences.
On nonbipartite lattices, the $d^1$ material does not host an $\mathrm{SU}(4)$ symmetry,
but still possesses a high symmetry enough to have interesting consequences.

Another important consequence of the emergent $\mathrm{SU}(4)$ symmetry is a
correspondence between spin and orbital degrees of freedom.  In quantum spin liquids
(QSLs), as it was most drastically demonstrated in Kitaev spin liquids, low-energy excitations
may be fractionalized into fermions~\cite{Kitaev2006}.  In the spin sector, the (fermionic)
spin-1/2 excitation is called spinon in distinction from magnon in the symmetry-broken
phase.  If there is an $\mathrm{SU}(4)$ symmetry in a system with fractionalized spin excitations,
there must be a fractionalized excitation even in the orbital sector.
We call this fermionic orbital excitation
\emph{orbitalon} in distinction from orbiton in the Jahn-Teller phase~\cite{Saitoh2001}.
While finding bosonic orbitons was one of the central topics in orbital physics~\cite{Tokura2000},
hunting fermionic orbitalons has just begun.  The $\mathrm{SU}(4)$ symmetry
must be an excellent guiding principle to search for fractionalization in the orbital sector.

We usually write down the $\mathrm{SU}(4)$ Heisenberg model
in the form of Eq.~\eqref{Eq.KK_SU4} in terms of the separate spin operators
$\bm{S}_j$ and orbital ones $\bm{T}_j$.
\begin{equation}
        H_\textrm{eff} = J \sum_{\langle ij \rangle} \Bigl(\bm{S}_i
\cdot \bm{S}_j+\frac{1}{4}\Bigr)\Bigl(\bm{T}_i \cdot
\bm{T}_j+\frac{1}{4}\Bigr), \label{Eq.KK_SU4}
\end{equation}
where $J>0$, $\bm{S}_j$, and $\bm{T}_j$ are (pseudo)spin-$1/2$
operators defined for each site $j$, and the sum is over nearest-neighbor
$ij$-bonds.  This is a special high-symmetry point of the Kugel-Khomskii
model~\cite{Kugel1982}.  A certain type of frustration involving
spin and orbital degrees of freedom exists in this Hamiltonian:  If the spin sector forms singlets,
the orbital sector forms triplets and \textit{vice versa}, so even a small
number of bonds have a strong frustration denying the singlet formation.
The frustration survives even on bipartite lattices, which allows us to
regard various lattices as candidate QSOLs.

We note that these highly symmetric $\mathrm{SU}(4)$ models are relevant to
materials other than $\alpha$-ZrCl$_3$ originally proposed in Ref.~\onlinecite{Yamada2018}.
For example, the relevance of an $\mathrm{SU}(4)$ QSOL has been discussed
for Ba$_3$CuSb$_2$O$_9$ (BCSO) with a decorated honeycomb
lattice structure~\cite{Zhou2011,Nakatsuji2012,Corboz2012}.
It turned out, however, that the estimated parameters for BCSO are
rather far from the model
with an exact $\mathrm{SU}(4)$ symmetry~\cite{Smerald2014}.
(See Refs.~\onlinecite{Ohkawa1983,Shiina1997,Wang2009,Kugel2015} for other proposed
realization of $\mathrm{SU}(4)$ symmetry, but they do not lead
to QSOL because of their crystal structures.)
The relevance of the $\mathrm{SU}(4)$ Heisenberg model has been discussed
beyond spin-orbital systems recently.
Especially, some of the two-dimensional (2D) systems with moir\'e superlattices
may be described by effective $\mathrm{SU}(4)$ models~\cite{Xu2018,Zhu2019}.

In this paper, we first introduce a notion of an emergent $\mathrm{SU}(4)$ symmetry in spin-orbital systems (Sec.~\ref{Sec:su4}),
derive it in the most general form, and discuss the possibility of various QSOLs
in the material realization of the $\mathrm{SU}(4)$ Heisenberg models (Sec.~\ref{Sec:emergent}).
Next, we consider the triangular lattice
as a representative nonbipartite lattice, discuss its realization, and introduce
an exotic frustrated Hamiltonian with an \textit{almost} $\mathrm{SU}(4)$ symmetry (Sec.~\ref{Sec:triangular}).
Finally, we will summarize this paper and remark some future directions (Sec.~\ref{Sec:discussions}).
To describe technical details, five Appendices~\ref{ticl}-\ref{trico} are given.

\section{$\mathrm{SU}(4)$ spin-orbital liquids}\label{Sec:su4}

\subsection{Dirac spin-orbital liquid}

Before moving on to the material proposal, we would like to review what
kind of spin-orbital liquids can be expected in $\mathrm{SU}(4)$ Heisenberg models.
The well-established and most famous one is a Dirac spin-orbital liquid in
the $\mathrm{SU}(4)$ Heisenberg model on the honeycomb lattice.
This state is found by a numerical study~\cite{Corboz2012}, but
is algebraically simple at the same time, so it is informative to
explain the analytic property of this ansatz state.

From variational Monte Carlo (VMC) and infinite projected entangled-pair state
(iPEPS) calculations, the $\mathrm{SU}(4)$ Heisenberg model
on the honeycomb lattice is expected to have a QSOL ground state~\cite{Corboz2012}.
The state is described by a $\pi$-flux Schwinger-Wigner ansatz of complex fermions
with an algebraic decay in correlation.

In order to derive the Schwinger-Wigner representation, first we rewrite
the Hamiltonian in terms of the $\mathrm{SU}(4)$ operators
up to a constant shift as
\begin{equation}
    H_\textrm{eff} = \frac{J}{4} \sum_{\langle ij \rangle} P_{ij} =
    \frac{J}{4} \sum_{\langle ij \rangle} \sum_{\alpha \beta} S_\alpha^\beta(i) S_\beta^\alpha(j),
\end{equation}
where a spin state at each site forms a fundamental representation
of $\mathrm{SU}(4)$, and we define $P_{ij}$ as the permutation operator
which swaps the states at sites $i$
and $j$.  $\mathrm{SU}(4)$ spin operators $S_\alpha^\beta(j)$ are obeying
\begin{equation}
    [S_\alpha^\beta,S_{\alpha^\prime}^{\beta^\prime}] = \delta_{\beta\alpha^\prime} S_\alpha^{\beta^\prime} - \delta_{\alpha\beta^\prime} S_{\alpha^\prime}^{\beta}. 
\end{equation}
Then, $S_\alpha^\beta(j)$ can be represented by
$S_\alpha^\beta(j)=f_{j\alpha}^\dagger f_{j\beta}$ using a complex fermion $f_{j\alpha}$
with $\alpha=1,\dots,4$.  This representation with a Gutzwiller projection
$\sum_\alpha f_{j\alpha}^\dagger f_{j\alpha} = 1$ will describe the $\mathrm{SU}(4)$
spin correctly.

\begin{figure}
\centering
\includegraphics[width=8.6cm]{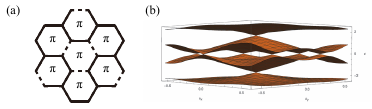}
\caption{(a) Gauge used in the $\pi$-flux mean-field solution on the honeycomb lattice.
For black solid bonds, $\eta_{ij} = 1$, while for black dashed bonds $\eta_{ij} = -1$.
(b) Dispersion of the $\pi$-flux mean-field solution on the honeycomb lattice.}
\label{dsol}
\end{figure}

After inserting this Schwinger-Wigner representation~\cite{Wen2002},
the mean-field Hamiltonian becomes
\begin{align}
    H_\textrm{MF}^{(1)} &= -\chi_0\sum_{\langle ij \rangle,\,\alpha} \eta_{ij} (f_{i\alpha}^\dagger f_{j\alpha} + h.c.),\label{MF1}
\end{align}
where $\eta_{ij} = \pm 1$ are determined as shown in Fig.~\ref{dsol}(a) and
$\chi_0$ is some constant. This choice of $\eta_{ij}$ corresponds to a $\pi$ flux through
every hexagonal plaquette. Eq.~\eqref{MF1} with a Gutzwiller projection
gives a variational wavefunction. The dispersion of this $\pi$-flux ansatz is
shown in Fig.~\ref{dsol}(b).  There are two degenerate Dirac cones at $\Gamma$
when it is quarter-filled.  Thus, this mean-field solution with a Gutzwiller projection
is a candidate Dirac spin-orbital liquid, where complex fermions are coupled to
some gauge field, with doubly degenerate Dirac cones.  This type of spin-orbital
liquids with an algebraic correlation is one typical QSOL expected in the $\mathrm{SU}(4)$
system.  This gapless property of the $\mathrm{SU}(4)$ Heisenberg model
on the honeycomb lattice is confirmed by various numerical techniques~\cite{Corboz2012}.

If we use the language of spin-orbital systems, the unbroken $\mathrm{SU}(4)$ symmetry
leads to two types of fractionalized excitations, spinons and orbitalons,
which are transformed to each other by the $\mathrm{SU}(4)$ rotation.
An unbiased density matrix renormalization group (DMRG) study also suggests
the existence of a symmetric Mott-insulating state in the large-$U$ limit of the
$\mathrm{SU}(4)$ Hubbard model~\cite{Zhu2019}.
We note that this $\pi$-flux ansatz with Dirac cones is analogous to
the Affleck-Marston approach~\cite{Affleck1988am,Corboz2012}.
However, it has recently been claimed that the original $\pi$-flux Dirac
spin-orbital liquid might be unstable with respect to the monopole perturbation,
leaving the question on the nature of the true ground state still open~\cite{Calvera2022}.

\subsection{Spinon-orbitalon Fermi surface}

Even within the Schwinger-Wigner representation, other phases of spinons and
orbitalons are possible depending on lattices and flux sectors.  A particularly interesting
case is the one with a Fermi surface formed by spinons and orbitalons where the
$\mathrm{SU}(4)$ symmetry is not broken.  This is a natural generalization of
the spinon Fermi surface theory to $\mathrm{SU}(4)$.

While a spinon-orbitalon Fermi surface is not expected on the honeycomb lattice,
it was demonstrated that it is a candidate ground state for the hyperhoneycomb lattice~\cite{Natori2018su4},
one of the best-known 3D generalizations of the honeycomb
lattice~\cite{Takayama2015}.

In the case of the hyperhoneycomb lattice, the following 0-flux mean-field Hamiltonian
is expected to describe the ground state.
\begin{align}
    H_\textrm{MF}^{(2)} &= -\chi_0^\prime \sum_{\langle ij \rangle,\,\alpha} (f_{i\alpha}^\dagger f_{j\alpha} + h.c.),
\end{align}
where $\chi_0^\prime$ is some constant. Interestingly, this state has a Fermi surface
at quarter filling (one fermion per site), so this mean-field solution describes the spinon-oribitalon Fermi
surface as long as the $\mathrm{SU}(4)$ symmetry is not broken.
An energetically unfavored $\pi$-flux state also possesses exotic Dirac
cones~\cite{Natori2018su4}, so the dynamics in the flux sector of the hyperhoneycomb
QSOL would also be interesting.

The Affleck-Marston-type flux state~\cite{Affleck1988am} may not be stabilized,
and may not be a good guess for the ground state away from half filling~\cite{Lieb1994}.
A further study is necessary to reveal the stability of spinon-orbitalon Fermi
surfaces more rigorously. We note that a Fermi surface is expected for the
$\mathrm{SU}(4)$ Heisenberg model on the triangular lattice~\cite{Keselman2020},
as well as the critical stripy state~\cite{Jin2021}.

\subsection{Majorana spin-orbital liquids}

Another possibility is a Majorana spin-orbital liquid with various (nodal) spectra.
Here we would not specify any mean-field solution and its spectrum because
we still do not know a lattice hosting such an exotic state.  However, the
$\mathrm{SO}(6)$ Majorana representation for $\mathrm{SU}(4)$ spins~\cite{Azaria1999,Wang2009}
to describe a Majorana spin-orbital liquid is mathematically fascinating, and thus
we would briefly review only the algebraic structure of this representation.
This Majorana representation is first proposed for the $\mathrm{SU}(4)$ Heisenberg model
on the square lattice~\cite{Wang2009}, but later it was found that the true ground state
may be a symmetry-broken phase~\cite{Corboz2011}.

Mathematically there is an accidental isomorphism between Lie algebras $\mathfrak{so}(6)$
and $\mathfrak{su}(4)$.
Strictly speaking, an accidental isomorphism can be used only for Lie algbebras, but
we abuse terminology like $\mathrm{SO}(6) \cong \mathrm{SU}(4)$, for simplicity.
Here, $\cong$ means local isomorphism.
Since $\mathrm{SU}(4) \cong \mathrm{SO}(6),$ we can also find an isomorphism between
an antisymmetric tensor representation of $\mathrm{SU}(4)$ and a vector representation
of $\mathrm{SO}(6).$  Although we will not explicitly demonstrate these isomorphisms,
this is the reason why we can construct an $\mathrm{SO}(6)$ Majorana representation.

The representation is similar to Kitaev's for the $\mathrm{SU}(2)$ spin~\cite{Kitaev2006}.
First, we divide the $\mathrm{SU}(4)$ fundamental representation into spin and orbital
degrees of freedom.  Then, a spin $\bm{S}_j$ and an orbital $\bm{T}_j$ can be decomposed
into a cross product of two sets of $\mathrm{SO}(3)$ Majorana fermions.
\begin{align}
    S_j^{\gamma} &= -\frac{i}{4} \varepsilon^{\alpha \beta \gamma} \eta_j^\alpha \eta_j^\beta, \\
    T_j^{\gamma} &= -\frac{i}{4} \varepsilon^{\alpha \beta \gamma} \theta_j^\alpha \theta_j^\beta,
\end{align}
where $\varepsilon^{\alpha \beta \gamma}$ is a Levi-Civita symbol, and $\bm{\eta}$ and $\bm{\theta}$ are $\mathrm{SO}(3)$ Majorana fermions with $\{\eta_i^\alpha, \eta_j^\beta\}=\{\theta_i^\alpha, \theta_j^\beta\} = 2 \delta_{ij}\delta^{\alpha \beta},$ and $\{\eta_i^\alpha, \theta_j^\beta\}=0.$
These 6 Majorana fermions per site provide a natural basis for the $\mathrm{SU}(4) \cong \mathrm{SO}(6)$ symmetry.
The Fock space is redundant and has a dimension $(\sqrt{2})^6= 8$ at each site.
Thus, we have to project it onto the 4-dimensional physical subspace in an
$\mathrm{SO}(6)$-symmetric way.

The simplest constraint for the projection would be
\begin{align}
    i\eta_j^x \eta_j^y \eta_j^z \theta_j^x \theta_j^y \theta_j^z = 1 \quad \textrm{for}\,\forall j, \label{case1}
\end{align}
or
\begin{align}
    i\eta_j^x \eta_j^y \eta_j^z \theta_j^x \theta_j^y \theta_j^z = -1 \quad \textrm{for}\,\forall j. \label{case2}
\end{align}
Indeed both Eq.~\eqref{case1} and Eq.~\eqref{case2} can simplify
the original $\mathrm{SU}(4)$ Hamiltonian and result in the same Majorana Hamiltonian.
In either case, all higher order terms in the $\mathrm{SU}(4)$ Heisenberg model
can be reduced into quartic terms:
\begin{align}
    H_\textrm{Majorana} \propto -\frac{1}{8}\sum_{\langle ij \rangle} \left( i\bm{\eta}_i \cdot \bm{\eta}_j + i\bm{\theta}_i \cdot \bm{\theta}_j \right)^2.
\end{align}
Thus, at a saddle point we can define a real mean field to solve self-consistent
equations: $\chi_{ij}^R = \langle i\bm{\eta}_i \cdot \bm{\eta}_j + i\bm{\theta}_i \cdot \bm{\theta}_j \rangle,$ and the mean-field Hamiltonian reads
\begin{align}
    H_\textrm{MF}^R = \sum_{\langle ij \rangle} \left[ -\frac{\chi_{ij}^R}{4} \left( i\bm{\eta}_i \cdot \bm{\eta}_j + i\bm{\theta}_i \cdot \bm{\theta}_j \right) + \frac{(\chi_{ij}^R)^2}{8} \right].
\end{align}
Notice that the mean field $\chi_{ij}^R = -\chi_{ji}^R$ is always real.

We note that the fermion number is not conserved except for the $Z_2$ parity, and
usually we make a mean-field ansatz wavefunction by filling a Fermi sea until half filling.
The projection onto the physical subspace is similar to that for
the Kitaev model~\cite{Kitaev2006}.  In this Majorana spin-orbital liquid,
spinons and orbitalons are in fact intertwined due to the
projection Eq.~\eqref{case1} or Eq.~\eqref{case2}, and thus we shall call them spin-orbitalons.

\section{Emergent $\mathrm{SU}(4)$ symmetry}\label{Sec:emergent}

\subsection{Honeycomb materials}

\begin{figure}
\centering
\includegraphics[width=8.6cm]{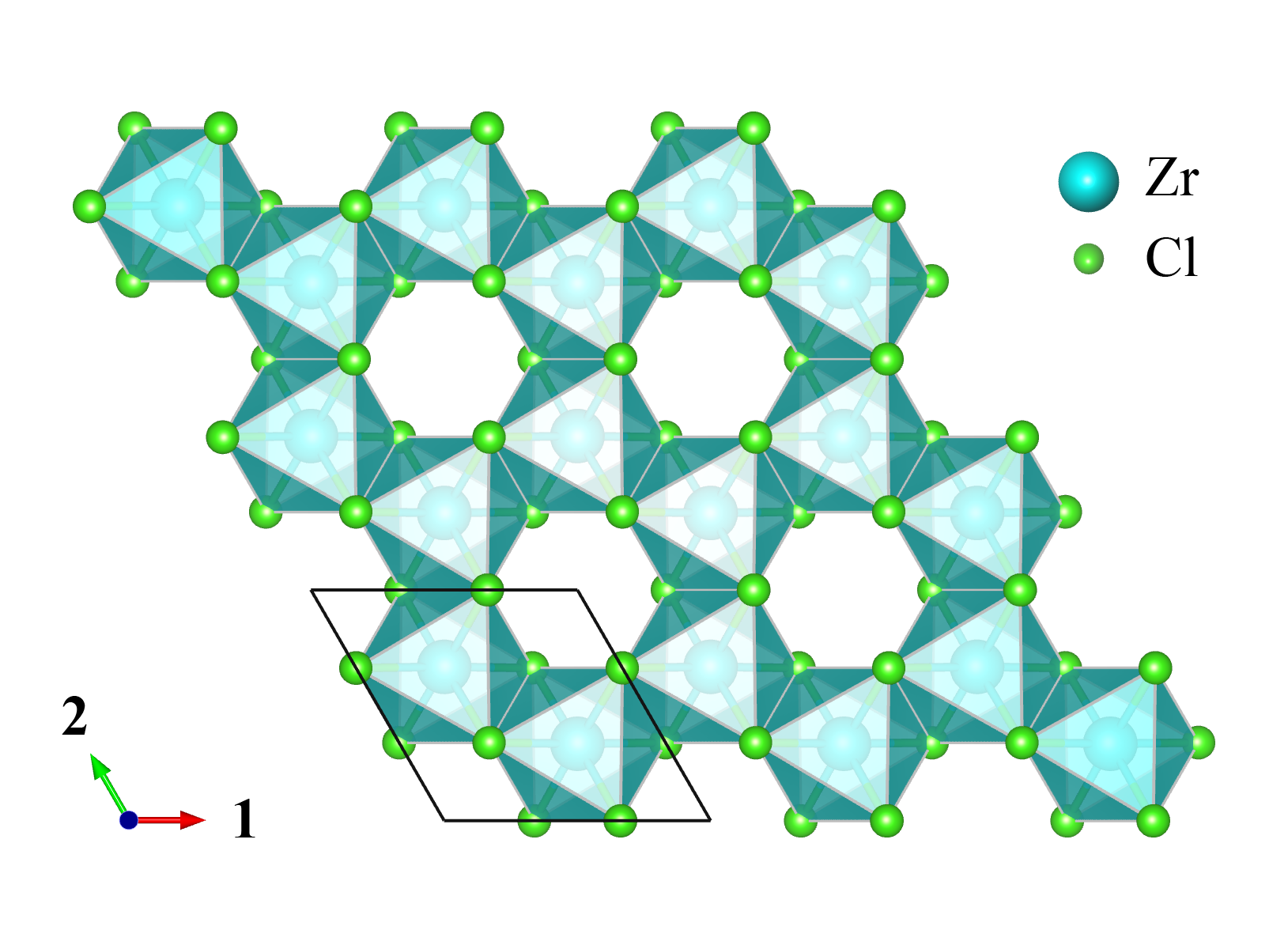}
\caption{Geometric structure of honeycomb $\alpha$-ZrCl$_3.$
Cyan and light green spheres represent Zr and Cl, respectively.
The crystallographic axes are shown and labelled as the 1- and 2-directions.
The figure is taken from Ref.~\onlinecite{Yamada2018}.}
\label{zrcl}
\end{figure}

From now on we will move on to the material side.
In many senses $\alpha$-ZrCl$_3$ is the first and most important
candidate for an emergent $\mathrm{SU}(4)$ symmetry.  This material
was reported in 1960s by Swaroop and Flengas~\cite{Swaroop1964Chem,Swaroop1964Phys}.
In the reported
structure, Zr$^{3+}$ is in the $d^1$ electronic configuration, octahedrally
surrounded by Cl$^{-}$.  The crystal structure is supposed to be honeycomb-layered
with a high symmetry~\cite{Swaroop1964Chem,Swaroop1964Phys} [see Fig.~\ref{zrcl}].
In the following discussions, we assume that $\alpha$-ZrCl$_3$ indeed forms
well-separated layers of the ideal honeycomb lattice.
It should be noted that, however, the crystal structure in
Refs.~\onlinecite{Swaroop1964Chem,Swaroop1964Phys} may be based on
a misaligned powder pattern~\cite{Daake1978}.
In addition, a recent density-functional theory calculation suggests
that this material might be susceptible to dimerization of the
honeycomb layers~\cite{Ushakov2020}.  If the crystal structure is in fact
different from the assumed honeycomb one, the theory should also be modified
accordingly.  Even if the crystal structure is modified, as long as the spin-orbit
coupling is unquenched, it probably leads to an exotic orbital magnetism.
On the other hand, we can replace atoms
as long as the $d^1$ electronic configuration is kept.  We can think of
$\alpha$-$MX_3$, with $M=$ Ti, Zr, Hf, \textit{etc.}, $X=$ F, Cl, Br, \textit{etc.}
They are also candidate materials to realize the $\mathrm{SU}(4)$ Heisenberg model
on the honeycomb lattice.
The case of $\alpha$-TiCl$_3$ is discussed separately in Appendix~\ref{ticl}.

The skeletal structure resembles that of $\alpha$-RuCl$_3$
which is known to be an important candidate for the Kitaev
honeycomb model~\cite{Plumb2014}.
We can regard $\alpha$-ZrCl$_3$ as a particle-hole inversion counterpart
of a transition metal halide $\alpha$-RuCl$_3$ because Ru$^{3+}$ has a
$d^5$ configuration, while Zr$^{3+}$ has a $d^1$ configuration.
The ground state is in the $J_\textrm{eff}=1/2$ subspace in the former,
whereas the ground state is in the $J_\textrm{eff}=3/2$ subspace in the latter.
We first demonstrate constructing $\mathrm{SU}(4)$ spin models for
an effective total angular momentum $J_\textrm{eff}=3/2$ on each $M$ of honeycomb
$\alpha$-$MX_3$, following Ref.~\onlinecite{Yamada2018}.

The $J_\textrm{eff}=3/2$ picture becomes asymptotically exact in the strong SOC limit.
This can be achieved by increasing the atomic number of $M$ from Ti to Hf.
The compounds $\alpha$-$M$Cl$_3$ with $M=$ Ti, Zr and related Na$_2$VO$_3$
have been already reported experimentally.
For simplicity, we only use
$\alpha$-ZrCl$_3$, although exactly the same discussion would apply
to $\alpha$-HfCl$_3,$ and other honeycomb systems $A_2M^\prime$O$_3$
($A=$ Na, Li, \textit{etc.}, $M^\prime=$ Nb, Ta, \textit{etc.}) as well.

\begin{figure}
\centering
\includegraphics[width=8.6cm]{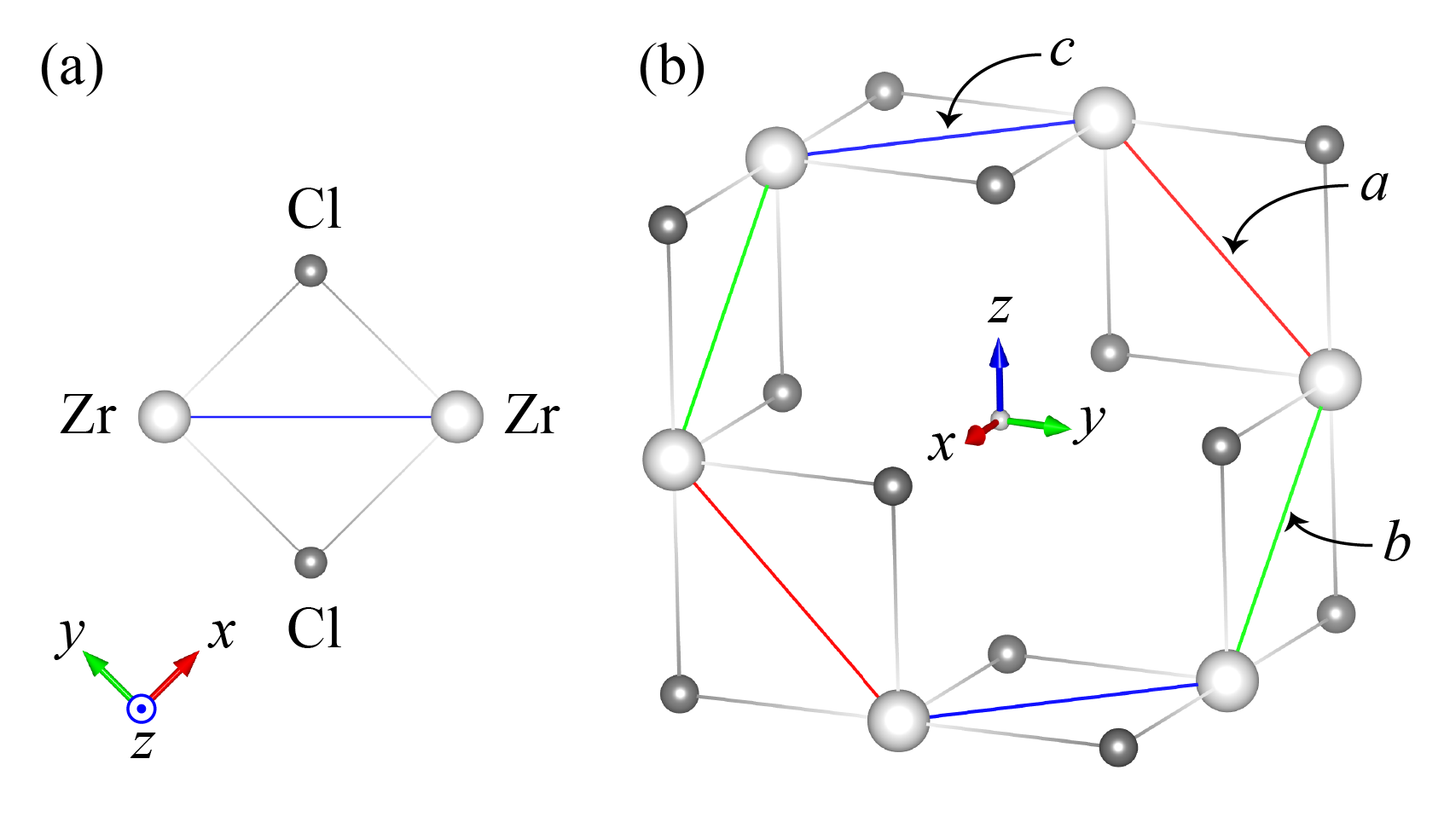}
\caption{(a) Superexchange pathways between two Zr ions connected by
a $c$-bond (blue) in $\alpha$-ZrCl$_3.$
White and grey spheres represent Zr and Cl atoms, respectively.
(b) Three different types of bonds in $\alpha$-ZrCl$_3.$
Red, light green, and blue bonds represent $a$-, $b$-, and $c$-bonds on the
$yz$-, $zx$-, and $xy$-planes, respectively.
The figure is taken from Ref.~\onlinecite{Yamada2018}.}
\label{honeycomb}
\end{figure}

\subsection{Effective Hamiltonian}\label{uij}

In the strong-ligand-field limit, the description with one electron
in the threefold degenerate $t_{2g}$-shell becomes accurate for $\alpha$-ZrCl$_3$.
The $t_{2g}$-orbitals ($d_{yz}$, $d_{zx}$, and $d_{xy}$-orbitals) are denoted
by $a$, $b$, $c$, respectively.
Let $a_{j\sigma}$, $b_{j\sigma}$, and $c_{j\sigma}$ represent annihilation operators
for these orbitals on the $j$th site of the honeycomb lattice with spin-$\sigma$, and
$n_{\xi\sigma j}$ with $\xi \in \{a,b,c\}$ be the corresponding number operators.
We also use this $(a,\,b,\,c)=(yz,\,zx,\,xy)$ notation for bonds:
each Zr --- Zr bond is labeled as $\xi$-bond ($\xi \in \{a,b,c\}$)
when the superexchange pathway is on the $\xi$-plane~\footnote{The Cartesian $xyz$ axes are defined as shown
in Fig.~\ref{honeycomb}(b).}, as depicted in Fig.~\ref{honeycomb}.

Although there are many ways to define a $J_\textrm{eff}=3/2$ spinor $\psi$,
we here use the following bases: \mbox{$\psi = (\psi_{\uparrow \uparrow},\psi_{\uparrow \downarrow},\psi_{\downarrow \uparrow},\psi_{\downarrow \downarrow})^t = (\psi_{3/2},\psi_{-3/2},\psi_{1/2},\psi_{-1/2})^t$}, where $\psi_{J^z}$ ($J^z=\pm 3/2,\, \pm 1/2$) is the annihilation operator for the $\ket{J=3/2, J^z}$ state.
Assuming the SOC is the largest electronic energy scale, except for the ligand field splitting,
fermionic operators can be projected onto the $J_\textrm{eff}=3/2$ states by inserting the quartet
$\psi_{j\tau\sigma}$ as follows.
\begin{align}
        a_{j\sigma}^\dagger &\to \frac{\sigma}{\sqrt{6}} (\psi_{j\uparrow \bar{\sigma}}^\dagger-\sqrt{3}\psi_{j\downarrow \sigma}^\dagger), \label{Eq.a} \\
        b_{j\sigma}^\dagger &\to \frac{i}{\sqrt{6}} (\psi_{j\uparrow \bar{\sigma}}^\dagger+\sqrt{3}\psi_{j\downarrow \sigma}^\dagger), \label{Eq.b} \\
        c_{j\sigma}^\dagger &\to \sqrt{\frac{2}{3}}\psi_{j\uparrow\sigma}^\dagger, \label{Eq.c}
\end{align}
where the indices $\tau$ and $\sigma$ of $\psi_{j\tau\sigma}$ represent the pseudoorbital
and pseudospin indices, respectively.  Here $\bar{\sigma}$ means an opposite spin to $\sigma.$
We begin from the following 6-component Hubbard Hamiltonian for $\alpha$-ZrCl$_3.$
\begin{align}
        H =& -t \sum_{\sigma, \langle ij \rangle \in \alpha} (\beta_{i\sigma}^\dagger \gamma_{j\sigma}+\gamma_{i\sigma}^\dagger \beta_{j\sigma})+ h.c. \nonumber \\
&+ \frac{U}{2} \sum_{j, (\delta,\sigma) \neq (\delta^\prime,\sigma^\prime)} n_{\delta\sigma j}n_{\delta^\prime \sigma^\prime j}, \label{Eq.original}
\end{align}
where $t$ is a real hopping parameter through the superexchange pathway shown in Fig.~\ref{honeycomb}(a), $U>0$ is the Hubbard term, $\langle ij \rangle \in \alpha$ means that the bond $\langle ij\rangle$ is
an $\alpha$-bond,
$\langle \alpha,\beta,\gamma \rangle$ runs over every cyclic permutation of $\langle a,b,c \rangle,$ and $\delta,\delta^\prime \in \{a,b,c\}.$
The effects of the Hund coupling $J_H$, not included explicitly in Eq.~\eqref{Eq.original},
are discussed in Appendix~\ref{hund}.
Simply by inserting Eqs.~\eqref{Eq.a}-\eqref{Eq.c}, we obtain
\begin{equation}
        H= -\frac{t}{\sqrt{3}} \sum_{\langle ij \rangle} \psi_i^\dagger U_{ij} \psi_j +h.c.
        + \frac{U}{2} \sum_{j} \psi_j^\dagger \psi_j (\psi_j^\dagger \psi_j-1), \label{Eq.Hub1}
\end{equation}
where $\psi_j$ is the aforementioned $J_\textrm{eff}=3/2$ spinor on the $j$th site,
and $U_{ij}=U_{ji}$ is a $4\times 4$ unitary matrix
\begin{equation}
        U_{ij} = \begin{cases}
    U^a = \tau^y \otimes I_2 & (\langle ij \rangle \in a) \\
    U^b = -\tau^x \otimes \sigma^z & (\langle ij \rangle \in b) \\
    U^c = -\tau^x \otimes \sigma^y & (\langle ij \rangle \in c)
  \end{cases},
\end{equation}
where $\bm{\tau}$ and $\bm{\sigma}$ are Pauli matrices
acting on the $\tau$ and $\sigma$ indices
of $\psi_{j\tau\sigma}$, respectively, and $I_N$ is an $N\times N$ identity matrix.
We note that $U^{a,b,c}$ are Hermitian, so $U_{ji}={U_{ij}}^\dagger
= U_{ij}$.

Now let us define an $\mathrm{SU}(4)$ gauge transformation,
\begin{equation}
        \psi_j \to g_j\cdot \psi_j, \qquad
        U_{ij} \to g_i U_{ij} g_j^\dagger,
\end{equation}
where $g_j$ is an element of $\mathrm{SU}(4)$ chosen for each site $j$.
For any loop $C$ on the honeycomb lattice, the $\mathrm{SU}(4)$ flux defined by a product $\prod_{\langle ij \rangle \in C} U_{ij}$ is invariant under the gauge transformation.

For each elementary hexagonal loop (which we call plaquette) $p$
in the honeycomb lattice with the coloring indicated
in Fig.~\ref{honeycomb}(b), the product becomes
\begin{equation}
        \prod_{\langle ij \rangle \in \hexagon_p} U_{ij}=U^a U^b U^c U^a U^b U^c=(U^a U^b U^c)^2 =-I_4,
\end{equation}
corresponding to an Abelian phase $\pi$.
Since all the loops in the honeycomb lattice are made of these plaquettes,
there exists an $\mathrm{SU}(4)$ gauge transformation which reduces
the model~\eqref{Eq.Hub1} to the $\pi$-flux
Hubbard model $H$ with a global $\mathrm{SU}(4)$ symmetry.
\begin{equation}
        H = -\frac{t}{\sqrt{3}} \sum_{\langle ij \rangle} \eta_{ij} \psi_i^{\dagger} \psi_j + h.c.
        + \frac{U}{2} \sum_{j} \psi_j^{\dagger} \psi_j (\psi_j^{\dagger} \psi_j -1), \label{Eq.piflux1}
\end{equation}
where the definition of $\eta_{ij}=\pm 1$, which is arranged
to insert a $\pi$ flux inside each plaquette, is shown in Fig.~\ref{dsol}(a).

At quarter filling, \textit{i.e.} one electron per site,
as is the case in $\alpha$-ZrCl$_3$, the ground state
becomes a Mott insulator for a sufficiently large $U/|t|$.
In this regime, the effective Hamiltonian for the spin and orbital degrees of
freedom, obtained by the second-order perturbation in $t/U$,
becomes the Kugel-Khomskii model exactly at the $\mathrm{SU}(4)$ point~\eqref{Eq.KK_SU4},
with $\bm{S}=\bm{\sigma}/2$, $\bm{T}=\bm{\tau}/2$, and $J=8t^2/(3U)$
in the basis set after the gauge transformation.
We note that the phase factor $\eta_{ij}$ cancels out in this second-order perturbation.
This $\mathrm{SU}(4)$ Heisenberg model on the honeycomb lattice is established
to host a gapless QSOL~\cite{Corboz2012}, so
we have found a possible realization of a Dirac spin-orbital liquid
in $\alpha$-ZrCl$_3$ with an \textit{emergent} $\mathrm{SU}(4)$ symmetry.

\subsection{Lieb-Schultz-Mattis-Affleck theorem}

The nontrivial property of this model
may be understood in terms of the Lieb-Schultz-Mattis-Affleck (LSMA) theorem for 
the $\mathrm{SU}(N)$ spin chains~\cite{LSM1961,Affleck1986,Lajko2017,YHO2018},
generalized to higher dimensions~\cite{LSM1961,Affleck1988,Oshikawa2000,Hastings2005,Totsuka2017}.
For the honeycomb lattice, which has two sites per unit cell, there is no LSMA constraint
for $\mathrm{SU}(2)$ spin systems~\cite{Jian2016}.
Nevertheless, as for the $\mathrm{SU}(4)$ spin system
we discuss in this paper, a two-fold ground-state degeneracy is
at least necessary to open a gap.
This implies the stability of a gapless QSOL phase observed in
the $\mathrm{SU}(4)$ Heisenberg model on the honeycomb lattice.

The claim of the LSMA theorem is as follows:
Under the unbroken $\mathrm{SU}(N)$ symmetry and translation symmetry,
the ground state of the $\mathrm{SU}(N)$ spin system with $n$ fundamental
representations per unit cell cannot be unique,
if there is a non-vanishing excitation gap and $n/N$ is fractional.
This rules out a possibility of a featureless Mott insulator phase,
which is defined as a gapped phase with a unique ground state
without any spontaneous symmetry breaking or topological order.

The original paper by Affleck and Lieb~\cite{Affleck1986} only
discussed one-dimensional (1D)
systems, so we would like to extend this theorem to higher dimensions
and systems with a space group symmetry.  The proof, based on Oshikawa's
flux insertion argument~\cite{Oshikawa2000}, is discussed in detail in
Appendix~\ref{lsma}.  The proof is not mathematically rigorous but physically intuitive.
Here we would just summarize the logic used in the proof.

In the $\mathrm{SU}(2)$ case, the inserted flux is
a magnetic flux constructed by $S^z$ operators, but in the $\mathrm{SU}(N)$ case we use
the following operator instead:
\begin{equation}
		I^0=\frac{1}{N} \begin{pmatrix}
1 & 0 & \cdots & 0 & 0 \\
0 & 1 &  & 0 & 0\\
\vdots &  & \ddots &  & \vdots \\
0 & 0 &  & 1 & 0 \\
0 & 0 & \cdots & 0 &-(N-1)
\end{pmatrix}.
\end{equation}
The diagonal elements obey $I^0 \mod 1 = 1/N$, so this changes the denominator
of the filling fraction from 2 to $N$.  This is the intuitive understanding of
the theorem, and would be applied to higher dimensions and the case with a
space group symmetry.

\subsection{Three-dimensional generalizations}

\begin{table}
        \centering
        \caption{\label{lattice}Tricoordinated lattices discussed in this paper.
Space groups are shown in number indices. Nonsymmorphic ones are underlined.
$n$ is the number of sites per unit cell.}
        \begin{tabular}{cccccc}
                Lattice name & $\mathrm{SU}(4)$ & \mbox{120\degree} bond & $n$ & Space group & LSMA \\
                \hline
                (10,3)-$a$ & \checkmark\footnotemark[1] & \checkmark & 4 & \underline{\textbf{214}} & \checkmark\footnotemark[2] \\
                (10,3)-$b$ & \checkmark\footnotemark[1] & \checkmark & 4 & \underline{\textbf{70}} & \checkmark\footnotemark[2] \\
                (10,3)-$c$ & $-$ & $-$ & 6 & \underline{\textbf{151}} & \checkmark \\
                (10,3)-$d$ & \checkmark\footnotemark[1] & $-$ & 8 & \underline{\textbf{52}} & \checkmark\footnotemark[2] \\
                (9,3)-$a$ & $-$ & $-$ & 12 & \textbf{166} & $-$ \\
                $8^2.10$-$a$ & \checkmark & \checkmark & 8 & \underline{\textbf{141}} & $-$ \\
				(8,3)-$b$ & \checkmark & \checkmark & 6 & \textbf{166} & \checkmark\footnotemark[3] \\
                stripyhoneycomb & \checkmark & \checkmark & 8 & \underline{\textbf{66}} & $-$ \\
                (6,3) & \checkmark & \checkmark & 2 & & \checkmark\footnotemark[4]
        \end{tabular}
\footnotetext[1]{The product of hopping matrices along every elementary loop is unity, 
resulting in the $\mathrm{SU}(4)$ Hubbard model with zero flux.}
\footnotetext[2]{Nonsymmorphic symmetries of the lattice are sufficient to protect a QSOL state,
hosting an crystalline spin-orbital liquid state [see Appendix~\ref{xsl}].}
\footnotetext[3]{Although the model has a $\pi$ flux, with an appropriate gauge choice the unit cell is not enlarged.  Therefore, the LSMA theorem straightforwardly applies to the $\pi$-flux $\mathrm{SU}(4)$ Hubbard model.}
\footnotetext[4]{While the standard LSMA theorem is not effective for
the $\pi$-flux $\mathrm{SU}(4)$ Hubbard model here,
the magnetic translation symmetry works to protect a QSOL state~\cite{LRO2017}.}
\end{table}

Generalized 3D honeycomb lattices are sometimes called tricoordinated lattices.
Recently, the classification of Kitaev spin liquids on various tricoordinated lattices
has been made~\cite{Hermanns2015Weyl,Hermanns2015BCS,Obrien2016}, so
we follow their strategy to extend the $\mathrm{SU}(4)$ physics to 3D.
We listed all the tricoordinated lattices considered in this paper on
Table~\ref{lattice}.  This table is based on Wells' classification of tricoordinated
lattices~\cite{Wells1977}.  We use a Schl\"afli symbol $(p,c)$ to label each lattice,
where $p$ is the shortest length of the elementary loops of the lattice,
and $c=3$ means the tricoordination of each vertex.
For instance, (6,3) is the 2D honeycomb lattice, and
all the other lattices are 3D lattices, distinguished by an additional letter
following Wells~\cite{Wells1977}.  $8^2.10$-$a$ is a nonuniform lattice and
the notation is different from the other lattices.

By generalizing the discussion of the honeycomb lattice to generic cases,
if the $\mathrm{SU}(4)$ orbital flux for
any loop $C$ is reduced to an Abelian phase $\zeta_C = \pm 1$, \textit{i.e.}
$    \prod_{\langle ij \rangle \in C} U_{ij} =
\zeta_C I_4 \quad (\textrm{for}\,^\forall C)$,
the Hubbard model will acquire the $\mathrm{SU}(4)$ symmetry.
This relation has been checked for each lattice in Table~\ref{lattice}.
We note that the flux inside is listed and included in Appendix~\ref{trico}.

A checkmark is put on the $\mathrm{SU}(4)$ column if the $\mathrm{SU}(4)$ symmetry exists.
Moreover, in order to form a stable structure,
the bonds from each site must form a 120-degree structure with an octahedral coordination.
This condition has again been checked for each lattice, and indicated
on the 120\degree~bond column~\cite{Obrien2016} of Table~\ref{lattice}.
Finally, we put a checkmark on the LSMA column
when the LSMA theorem implies the existence of ground state degeneracy or
gapless excitations for the resulting $\mathrm{SU}(4)$ Hubbard model.
For example, the LSMA theorem is applicable to
the (8,3)-$b$ lattice because $n/N=6/4$ is fractional.

We note that a 3D version of Na$_2$VO$_3$ has already been reported~\cite{Rudorff1956}.
Therefore we can  expect synthesis of various 3D polymorphs
of ZrCl$_3$ or $A_2M^\prime$O$_3$ with $A=$ Na, Li and $M^\prime=$ Nb, Ta, similarly
to 3D $\beta$-Li$_2$IrO$_3$~\cite{Takayama2015} and $\gamma$-Li$_2$IrO$_3$~\cite{Modic2014}.

\section{Triangular $d^1$ system with a broken $\mathrm{SU}(4)$ symmetry}\label{Sec:triangular}

\begin{figure}
\centering
\includegraphics[width=6cm]{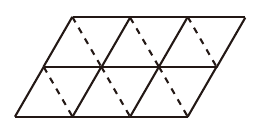}
\caption{Triangluar $d^1$ model.  Solid bonds have the $\mathrm{SU}(4)$ Heisenberg
interaction, but dashed bonds have an exotic interaction Eq.~\eqref{Eq.exotic}.
If we ignore dashed bonds, it becomes the $\mathrm{SU}(4)$
Heisenberg model on the square lattice~\cite{Corboz2011}.
}
\label{tri}
\end{figure}

It would be interesting to investigate $\mathrm{SU}(4)$
Heisenberg models on nontricoordinated lattices.  Especially, on the
lattice with 1 or 3 sites per unit cell, the LSMA theorem can
exclude the possibility of a simply gapped $Z_2$ spin liquid
and suggests a $Z_4$ QSOL or unusual SET phases instead.
This can be understood by applying
the proof of the LSMA theorem to a cylinder boundary condition
because the fourfold ground state degeneracy on a cylinder suggests the existence of a
gapless edge mode, or a topological order beyond $Z_2$ topological order,
for example.
The case of the triangular lattice is also mentioned in Ref.~\onlinecite{Natori2018su4}.

From now on, we only consider a triangular lattice case for simplicity.
Moreover, it may be relevant to some accumulated
graphene/transition metal dichalcogenide (TMDC) systems~\cite{Constantin2019}.
We can easily expect the existence of a $\mathrm{U}(1)$ spin liquid state even for
the $\mathrm{SU}(4)$ Heisenberg model on the triangular lattice~\cite{Keselman2020}.
However, unfortunately
real triangluar $d^1$ systems cannot host an exact $\mathrm{SU}(4)$ Heisenberg model.
Instead, as we will show in the following, we find a $\Gamma^5$ flux inside
each triangluar plaquette and the resulting spin-orbital model becomes exotic,
reflecting this additional (non-Abelian) flux.

Similarly to Ba$_3$IrTi$_2$O$_9$~\cite{Catuneanu2015}, which is a triangular $d^5$
Kitaev material, we can imagine a triangular $d^1$ system as a starting point.
In this case, each triangular plaquette binds the following flux:
\begin{align}
    \prod_{\langle ij \rangle \in \triangle} U_{ij}=U^a U^b U^c =: i\Gamma^5.
\end{align}
For simplicity, we use a chiral representation as follows:
\begin{align}
    \Gamma^5=-\tau^z \otimes I_2 =
    \begin{pmatrix}
        -I_2 & 0 \\
        0 & I_2
    \end{pmatrix}.
\end{align}

A gauge transformation can always concentrate a flux matrix to only one bond
for each triangular plaquette, so it is enough to focus on one bond $\langle ij \rangle$ with
$U_{ij} = i\Gamma^5$ in order to derive an effective spin-orbital model by the second-order
perturbation in $t/U.$  The rest of the bonds are all $\mathrm{SU}(4)$-symmetric, in which case
the discussion is completely parallel to the honeycomb case.  As for a bond with
$U_{ij} = i\Gamma^5,$ the second-order perturbation leads to the following spin-orbital
model:
\begin{equation}
    H_{ij} = J\Bigl(\bm{S}_i \cdot \bm{S}_j+\frac{1}{4}\Bigr)\Bigl(T_i^z T_j^z-T_i^x T_j^x-T_i^y T_j^y+\frac{1}{4}\Bigr), \label{Eq.exotic}
\end{equation}
if $\langle ij \rangle$ is a dashed bond shown in Fig~\ref{tri}.
We can expect an exotic frustration, which is different from that in
the $\mathrm{SU}(N)$ Heisenberg model.  To the best of our knowledge, there is
no previous study for this model, so it is worthwhile to study it here.

\begin{figure}
\centering
\includegraphics[width=8.6cm]{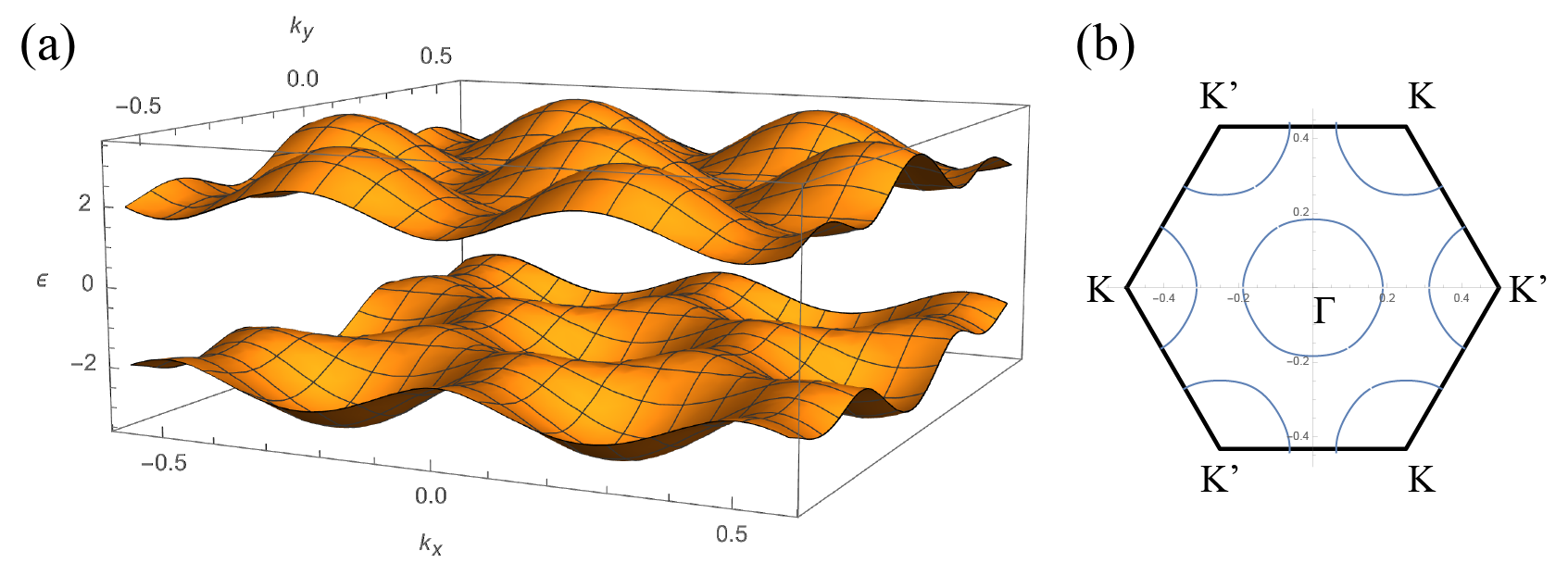}
\caption{(a) Band structure of the $\Gamma^5$-flux state.  All bands are doubly
degenerate due to the time-reversal and inversion symmetries.
(b) Fermi surfaces (blue lines) of the $\Gamma^5$-flux state at quarter filling.
The Brillouin zone is shown by black lines.}
\label{gamma5_fs}
\end{figure}

Then, what kind of QSOLs are relevant to this exotic model?
One of the most natural possibilities is the $\Gamma^5$-flux state.  This state
is described by the following trial wavefunction $\ket{\Psi_\textrm{GS}}$.
\begin{equation}
\ket{\Psi_\textrm{GS}}=P_\textrm{GW}\ket{\Psi_\textrm{free}},
\end{equation}
where $\ket{\Psi_\textrm{free}}$ is the free-fermionic ground state of the above
model with the $\Gamma^5$ flux in the case of $U=0$ at quarter filling, and $P_\textrm{GW}$ is the
Gutzwiller projection onto the space with $N_j=1$ for each $j$.
The correlation effect of $U\to\infty$ is included in the Gutzwiller projection.
Indeed, this state has a spinon-oribitalon Fermi surface.
As shown in Fig.~\ref{gamma5_fs}, two degenerate bands cross the Fermi level at
quarter filling and the cross section consists of circular Fermi surfaces.

However, this state most probably suffers from the
Bardeen-Cooper-Schrieffer (BCS) instability~\cite{Hermanns2015BCS}.
The twofold degeneracy of bands and the almost isotropic Fermi surface allow
the following BCS ground state instead of the original wavefunction.
\begin{align}
\ket{\Psi_\textrm{GS}^\prime} &=P_\textrm{GW}\ket{\Psi_\textrm{BCS}}. \\
\ket{\Psi_\textrm{BCS}} &= \prod_{\bm{k}} \left(u_{\bm{k}} + v_{\bm{k}} f_{-\bm{k}\downarrow}^\dagger f_{\bm{k}\uparrow}^\dagger \right)\ket{\Psi_\textrm{free}},
\end{align}
where the product about $\bm{k}$ is taken over the Fermi surface,
$u_{\bm{k}}$ and $v_{\bm{k}}$ are variational parameters with $u_{\bm{k}}^2 + v_{\bm{k}}^2 = 1$,
and $f_{\bm{k}\sigma}^\dagger$ is a creation operator of a spinon/orbitalon with
a momentum $\bm{k}$, where $\sigma = \uparrow,\downarrow$ labels the pseudospin
index of the Kramers band degeneracy.  This describes the standard s-wave
pairing of the Cooper pair, while other pairings are also possible.

The energy of the proposed state $\ket{\Psi_\textrm{GS}^\prime}$
cannot easily be evaluated and probably it requires a VMC simulation about
$u_{\bm{k}}$ and $v_{\bm{k}}$.  This state describes a kind of gapped spin liquids,
while its property is still obscure.  Whether or not this state is stabilized is
determined from the comparison of energy with other candidate states.
The energetic comparison of candidate states based on VMC is left for the future work.

Discussions here are relevant to 1T-TaS$_2$~\cite{Law2017,Yu2017,Murayama2020}
in a symmetric phase without a structural distortion.  However, the so-called
Star-of-David structure appears after the charge density wave transition,
which destroys the orbital degeneracy of the $J_\textrm{eff}=3/2$ states.
If the symmetric
phase survives at very low temperature, 1T-TaS$_2$ should also be an
important playground for the quasi-$\mathrm{SU}(4)$ magnetism.

NaZrO$_2$ is also a candidate for the same triangular $d^1$ state,
though the density functional theory (DFT) claims that it is in
a nonmagnetic metallic state~\cite{Assadi2018}.  It could possibly
lead to the above model after the Mott transition.
A DFT study for LiZrO$_2$ was also found~\cite{Singh2004}.

\section{Discussion} \label{Sec:discussions}

In this paper, we made a comprehensive study on various $d^1$ spin-orbit coupled systems and
discovered that the $\mathrm{SU}(4)$ Heisenberg models appear generically on many
tricoordinated bipartite lattices.
A part of the results presented in this work were already
announced in the previous short communication~\cite{Yamada2018}.
Expanding the original Letter~\cite{Yamada2018},
in this paper we have presented (i) the proof of the LSMA theorem generalized to
higher dimensions, (ii) discussions on the triangular lattice $d^1$ system,
and (iii) the flux structure of various tricoordinated bipartite lattices.

Even on nonbipartite lattices like the triangular
lattice, the $d^1$ model is exotic and worth investigating, while they do not host
a complete $\mathrm{SU}(4)$ symmetry.
The study of actual ground states for those models is left for future work,
though we expect QSOLs in general, possibly described by Dirac spin-orbital liquid,
spinon-orbitalon Fermi surface liquid, or more exotic Majorana liquids.

The Jahn-Teller term which couples the orbital to the lattice has not been discussed.
It typically breaks a symmetry of the lattice, resulting in a Jahn-Teller transition
to the low-symmetry phase~\cite{Tokura2000}.  In order for the symmetric phase to survive,
the itinerant quantum fluctuation which can tunnel between classical ground states
may be necessary.
Thus, the competition between QSOLs and Jahn-Teller phases (orbital order) can be understood
in terms of the spinon/orbitalon band width $W \sim J = 8t^2/(3U)$~\cite{Khaliullin2000}.
If $J$ is large enough compared to the phonon energy scale to stabilize
the (orbital) symmetric state, then the kinetic energy gain of orbitalons may destabilize
the Jahn-Teller order.  Thus, such energy gain may be maximized around the Mott transition, and
thus the 4$d$- or 5$d$-materials with a smaller $U$ may be beneficial.

An indirect sign of a realization of QSOL state in real materials would be the
absence of long range order down to the lowest temperatures.
Experimentally, muon spin resonance ($\mu$SR) or nuclear magnetic resonance (NMR)
experiments can rule out the existence of long-range magnetic ordering or spin freezing
in the spin sector. In the orbital sector, a possible experimental signature
to observe the absence of orbital ordering or freezing should be
electron spin resonance (ESR)~\cite{Han2015} or
extended X-ray absorption fine structure (EXAFS)~\cite{Nakatsuji2012}.
Especially, (finite-frequency) ESR can observe the dynamical Jahn-Teller effect~\cite{Nasu2013,Nasu2015},
where the $g$-factor isotropy directly signals the quantum fluctuation between
different orbitals~\cite{Han2015,Bersuker1975,Bible1970},
\textit{i.e.} the $\mathrm{SU}(2)$ subgroup symmetry in the orbital sector may be evident in the $g$-factor isotropy.
This is also applicable to our $t_{2g}$ case because of the shape difference in the $J_\textrm{eff}=3/2$
orbitals~\cite{Romhanyi2017}, and the static Jahn-Teller distortion will result in
the anisotropy in the in-plane $g$-factors~\cite{Iwahara2017}.  Here we note that
the trigonal distortion existing \textit{a priori} in real materials only splits the degeneracy
between the out-of-plane and in-plane $g$-factors, and the splitting of the two
in-plane modes clearly indicates some (\textit{e.g.} tetragonal) distortion.

The emergent $\mathrm{SU}(4)$ symmetry would result in coincidence between the
time scales of two different excitations for spins and orbitals observed
by NMR and ESR, respectively.

On the other hand, the direct detection of orbitalons may be more challenging.
Orbitalons carry an orbital angular momentum.
Magnetically an orbital angular momentum is indistinguishable and mixed
with a spin by SOC.  However, since the orbital fluctuation is coupled to the lattice,
an electric field, light, or X-rays can directly affect the orbital sector~\cite{Tokura2000}.
Especially, a light beam with an orbital angular momentum has been investigated
recently~\cite{Marrucci2006}, and may be useful for the detection of orbitalons.
It will be an interesting problem to discover the connection between such technology and fractionalized
orbital excitations.

Such orbital physics can be sought in other systems like $f$-electron systems.
For example, ErCl$_3$ may have twofold orbital degeneracy at low temperature~\cite{Kramer1999,Kramer2000}.
In many cases, orbitals have twofold degeneracy at most, so the highest achievable symmetry
of QSOLs in spin-orbital materials is $\mathrm{SU}(4).$
Whether it is possible to realize $\mathrm{SU}(6)$ spin systems in spin-orbital
systems is an interesting open question.  So far a cold atomic system is the only
candidate for $\mathrm{SU}(6)$~\cite{Nataf2016}.
The exploration of hitherto unknown materials with exotic symmetries
is still far from being finished, and it is a future problem to make a catalog of these systems.

\begin{acknowledgments}
We thank V.~Dwivedi, M.~Hermanns, H.~Katsura, K.~Kitagawa,
M.~Lajk\'o, F.~Mila, S.~Nakatsuji, K.~Shtengel, Y.~Tada, S.~Tsuneyuki, and
especially I.~Kimchi for helpful comments.
The crystal data have been taken from Materials Project~\cite{Jain2013}, drawn by VESTA~\cite{Momma2011}.
M.G.Y. is supported by the Materials Education program for the future leaders in Research, Industry, and Technology (MERIT), and by JSPS.
M.G.Y. is also supported by Multidisciplinary Research Laboratory System for Future Developments, Osaka University.
This work was supported by JST CREST Grant Numbers JPMJCR19T2 and JPMJCR19T5, Japan,
by JSPS KAKENHI Grant Numbers JP15H02113, JP17J05736, and JP18H03686, and by JSPS Strategic International Networks Program No. R2604 ``TopoNet''.
We acknowledge the support of the Max-Planck-UBC-UTokyo Centre for Quantum Materials.
This research was supported in part by the National Science Foundation under Grant No. NSF PHY-1748958.
\end{acknowledgments}

\appendix

\section{$\alpha$-TiCl$_3$}\label{ticl}

As for $\alpha$-TiCl$_3$, a structural transition and opening
of a spin gap at $T=217$ K have been reported~\cite{Ogawa1960}.
This implies a small SOC, as it is consistent with
a massively degenerate manifold of spin-singlets
expected in the limit of a vanishing SOC~\cite{Jackeli2007}.

We try to capture the physics of $\alpha$-TiCl$_3$ by the model
without SOC.  The model itself was already discussed in Section~VIB
of Ref.~\onlinecite{Normand2008} and Section~IIIA of Ref.~\onlinecite{Chaloupka2011}.
This weak-SOC limit is interesting as the valence bond liquid-type states are
expected and would potentially explain the observed spin gap behavior.

In addition to the above references, we would like to give an insight
from the $\mathrm{SU}(4)$ symmetry.  Indeed, the model at $J_H=0$
is ``locally'' $\mathrm{SU}(4)$-symmetric when we flip active orbital
on one of the two sites on an isolated bond.  Thus, locally the spin-singlet
orbital-triplet state, or the spin-triplet orbital-singlet state will lower
the energy, potentially leading to the resonating valence bond-like state
by covering the honeycomb lattice by $\mathrm{SU}(4)$ dimers.

The above picture is very naive but potentially explains the valence bond
formation accompanied by the spin gap transition from the $\mathrm{SU}(4)$
viewpoint.  Though there is no global $\mathrm{SU}(4)$ symmetry in the weak-SOC
limit, the local $\mathrm{SU}(4)$ symmetry is still useful and is worth mentioning
in this Appendix.

\section{$\mathrm{SU}(4)$-breaking terms}\label{hund}

Of course, real materials do not have a complete $\mathrm{SU}(4)$ symmetry
and we have to think of the effect of $\mathrm{SU}(4)$-breaking terms on the
spin-orbital liquid states.  Especially, we consider the case of $\alpha$-ZrCl$_3$
and discuss what kind of $\mathrm{SU}(4)$-breaking terms may exist.

The most relevant $\mathrm{SU}(4)$-breaking term would be the Hund coupling $J_H$.
The Hamiltonian can be written in the simplest form~\cite{Georges2013,Yamada2018} as
\begin{widetext}
\begin{align}
    H =& -t \sum_{\sigma, \langle ij \rangle \in \alpha} (\beta_{i\sigma}^\dagger \gamma_{j\sigma}+\gamma_{i\sigma}^\dagger \beta_{j\sigma})+ h.c. + \sum_{j} \left[ \frac{U-3J_H}{2} N_j(N_j-1) -2J_H \bm{s}_j^2 -\frac{J_H}{2} \bm{L}_j^2 +\frac{5}{2}J_H N_j \right],
\end{align}
\end{widetext}
where $\alpha$, $\beta$, and $\gamma$ are defined in the same way as
Eq.~\eqref{Eq.original}, $N_j$ is a number operator, $\bm{s}_j$ is a total spin,
and $\bm{L}_j$ is a total effective angular momentum within the $t_{2g}$ manifold.
It is easy to see that the perturbation from the original Hamiltonian
(Eq.~\eqref{Eq.original}) is small when $J_H/U \sim \mathcal{O}(0.1)$,
as long as the total $N_j$ is conserved.

In addition, it is not difficult to show that in the second-order perturbation the
contribution breaking the original $\mathrm{SU}(4)$ symmetry always involves an
virtual state with an energy higher than the lowest order by $\lambda$ or $J_H.$
Anyway, we can conclude that, as long as we ignore higher order contributions of
$J_H/U \sim \mathcal{O}(0.1)$, the emergent $\mathrm{SU}(4)$ symmetry would be robust.

We note that recently it was argued that $\mathcal{O}(0.1)$ perturbation
of $J_H/U$ would not destabilize the $\mathrm{SU}(4)$ spin liquid
in the case of BCSO~\cite{Natori2019}.  Although it is not clear this result is applicable to
$\alpha$-ZrCl$_3,$ we can expect that the stability region of a size $\mathcal{O}(0.1)$
will be reproduced for $\alpha$-ZrCl$_3,$ too, by similar mean-field and variational
calculations.  While this is a preliminary discussion, further studies will disclose
the effects of $J_H/U$ in the future.

\section{Crystalline spin-orbital liquids}\label{xsl}

Crystalline spin liquids (XSL)~\cite{Yamada2017XSL} are defined
originally for Kitaev models and the discussion is in Ref.~\onlinecite{Yamada2017XSL}.
We would quickly review the definition and generalize this notion to
$\mathrm{SU}(4)$-symmetric models based on the Lieb-Schultz-Mattis-Affleck (LSMA) theorem.

In the context of gapless Kitaev spin liquids as originally proposed in
Ref.~\onlinecite{Yamada2017XSL}, a crystalline spin liquid is
defined as a spin liquid state where a gapless point (or a gapped
topological phase) is protected not just by the unbroken time-reversal
or translation symmetry, but by the space group symmetry of the lattice.
This is a simple analogy with a topological crystalline insulator, where
a symmetry-protected topological order is protected by some space group symmetry.

Differently from topological crystalline insulators, the classification or identification
of crystalline spin liquids is not easy.  This is because a symmetry could be
implemented \textit{projectively} in spin liquids and the representation of the symmetry (action)
becomes a projective (fractionalized) one.  The classification depends not only on
its original symmetry of the lattice but also on its PSG, so there are a macroscopic
number of possible crystalline spin liquids.  The only thing we can do is to identify
the mechanism of the symmetry protection for each specific case.
In Ref.~\onlinecite{Yamada2017XSL}, two Kitaev spin liquids are identified, one with
three-dimensional (3D) Dirac cones, and the other with a nodal line protected by the
lattice symmetry, not by the time-reversal symmetry~\cite{Obrien2016}.

Sometimes, however, extended Lieb-Schultz-Mattis-type (LSM-type)
theorems can prove the existence of a gapless point
or a topological state in the gapped case.  Thus, the LSM theorem can potentially
prove that some spin liquid is XSL without a microscopic investigation,
if we ignore whether it is gapped or gapless~\cite{WPVZ2015}.
This is a subtle point, but LSM-type theorems
extended to include a nonsymmorphic symmetry is very powerful to discuss the
property of spin liquids abstractly.

Next, we would like to discuss the generalization of the concept of
XSL to $\mathrm{SU}(4)$-symmetric models.
In the (10,3) lattices listed in Table~\ref{lattice}, the unit cell
consists of a multiple of 4 sites, and thus the generalized LSMA theorem
seems to allow a featureless insulator if we only consider the translation.
Following Refs.~\onlinecite{PTAV2013,WPVZ2015,PWJZ2017}, however,
we can effectively reduce the size
of the unit cell by dividing the unit cell by the nonsymmorphic
symmetry, and thus the filling constraint becomes tighter with a
nonsymmorphic space group.  Even in the (10,3) lattices, the gapless
QSOL state can be protected by the further extension of the LSMA
theorem. We call them crystalline spin-orbital liquids
(XSOLs) in the sense that these exotic phases are protected
in the presence of both the $\mathrm{SU}(4)$
symmetry and (nonsymmorphic) space group symmetries.

\section{Details of the Lieb-Schultz-Mattis-Affleck theorem}\label{lsma}

The $\mathrm{SU}(N)$ Heisenberg model on the two-dimensional (2D) honeycomb
lattice admits the application of the LSMA
theorem~\cite{LSM1961,Affleck1986,Oshikawa2000,Hastings2005} for $N>2$.
However, the original paper by Affleck and Lieb~\cite{Affleck1986} only
discussed one-dimensional (1D)
systems, so we would like to extend the claim to higher dimensions
and systems with a space group symmetry.
Let us first consider a periodic 2D lattice with the
primitive lattice vectors $\bm{a}_{1,2}$,
as defined in Fig.~\ref{zrcl} in the main text.
We define the lattice translation operators $\trans_\mu$
along $\bm{a}_\mu$ for $\mu=1,2$.

Here we consider the case with a fundamental representation on each
site of the honeycomb lattice, which includes the $\mathrm{SU}(4)$ Heisenberg
model discussed in the main text.
We call each basis of the $\mathrm{SU}(N)$ fundamental representation \textit{flavor}.
The Hamiltonian of the $\mathrm{SU}(N)$ Heisenberg model on the honeycomb lattice in general can
be written as
\begin{equation}
		H_{\mathrm{SU}(N)}=\frac{J_a}{N} \sum_{\langle ij\rangle \in a} P_{ij} +\frac{J_b}{N} \sum_{\langle ij\rangle \in b} P_{ij} +\frac{J_c}{N} \sum_{\langle ij\rangle \in c} P_{ij}, \label{eq.Heisenberg}
\end{equation}
up to constant terms, where $J_\gamma$s are the bond-dependent coupling constants for the
$\gamma$-bonds, as defined in the main text, and $P_{ij}$ is the permutation operator
of the flavors between the $i$th and $j$th sites.  The translation symmetries,
$\trans_1$ and $\trans_2,$ exist independently of the values of $J_\gamma$s, so
the following discussions apply to any positive $J_\gamma$s.
Since the spin-1/2 Heisenberg antiferromagnetic interaction for the
$\mathrm{SU}(2)$ spin can also be written as Eq.~\eqref{eq.Heisenberg} with $N=2$
dimensional Hilbert space at each site.

Now we discuss the generalization of the LSMA theorem to $\mathrm{SU}(N)$ spin
systems~\cite{Affleck1986,Lajko2017,Totsuka2017}
in 2 dimensions following the logic of Ref.~\onlinecite{Oshikawa2000}.
One of the generators $I^0$ of the $\mathrm{SU}(N)$ in the fundamental
representation is given by the traceless $N\times N$
diagonal matrix:
\begin{equation}
		I^0=\frac{1}{N} \begin{pmatrix}
1 & 0 & \cdots & 0 & 0 \\
0 & 1 &  & 0 & 0\\
\vdots &  & \ddots &  & \vdots \\
0 & 0 &  & 1 & 0 \\
0 & 0 & \cdots & 0 &-(N-1)
\end{pmatrix}.
\end{equation}
We introduce an Abelian gauge field $\bm{\calA} (\bm{r})$,
which couples to the charge $I^0$, where
$\bm{r}$ is the coordinate.

We assume that the (possibly degenerate) ground
states are separated from the continuum of the excited states
by a nonvanishing gap, and that the gap does not collapse during
the flux insertion process discussed below.
We consider the system consisting of $L_1 \times L_2$ unit cells
on a torus, namely with periodic boundary conditions
$\bm{r} \sim \bm{r} + L_1 \bm{a}_1 \sim \bm{r} + L_2 \bm{a}_2$.
A ground state, which is $\mathrm{SU}(N)$-symmetric and 
has a definite crystal momentum
(\textit{i.e.} eigenstate of $\trans_{\mu}$ with $\mu=1,\,2$),
is chosen as the initial state.
We adiabatically increase the gauge field from $\bm{\calA}=0$
to $\bm{\calA} = \bm{k}_1/L_1$, so that
the ``magnetic flux'' contained in the ``hole'' of the torus
increases. When the ``magnetic flux'' reaches the unit flux quantum $2\pi,$
the Hamiltonian of the system becomes equivalent to the initial one.
This happens precisely when the Hamiltonian is obtained from
the original Hamiltonian with a large gauge transformation.
The minimal large gauge transformation with respect to the charge
$I^0$ is given by
\begin{equation}
\calU_1 = \exp{\left[
\frac{i}{L_1} \sum_{\bm{r}} \bm{k}_1 \cdot \bm{r} I^0(\bm{r})
\right]},
\end{equation}
where $\bm{k}_\mu$s are primitive reciprocal lattice vectors satisfying
\begin{equation}
 \bm{k}_\mu \cdot \bm{a}_\nu = 2 \pi \delta_{\mu \nu}.
\end{equation}

The large gauge transformation satisfies the commutation relation,
\begin{equation}
 \calU_1 \trans_1   = \trans_1 \calU_1
\exp{\left[
		\frac{2\pi i}{L_1} \Bigl( I^0_T -
\sum_{\bm{r}\cdot\bm{k}_1 = 2\pi (L_1 -1) } L_1 I^0\left( \bm{r} \right) \Bigr)
 \right]} .
\end{equation}
Here $I^0_T = \sum_{\bm{r}} I^0(\bm{r})$.
Since the ground state is assumed to be an $\mathrm{SU}(N)$-singlet when the number of sites
is a multiple of $N,$
it belongs to the eigenstate with $I^0_T = 0.$
Furthermore, because eigenvalues of $I^0(\bm{r})$ are equivalent to $1/N \mod{1},$ we find,
\begin{equation}
 {\trans_1}^{-1} \calU_1 \trans_1  \sim \calU_1 e^{-(2\pi i n L_2/N)},
\end{equation}
where $n$ is the number of sites in the unit cell.

Since the uniform increase in the vector potential does not
change the crystal momentum, this phase factor due to the
large gauge transformation alone gives the
change of the crystal momentum in the flux insertion process.
Choosing $L_2$ to be coprime with $N,$ we find 
a nontrivial phase factor when $n/N$ is not an integer.
This implies that, if $n$ is not an integer multiple of $N$,
the system must be gapless or has degenerate ground states.

For the honeycomb lattice, $n=2,$ and there is no LSMA constraint
for $\mathrm{SU}(2)$ spin systems. In contrast, for the $\mathrm{SU}(4)$ spin system
we discussed in the main text, the ground-state degeneracy (or gapless
excitations) is required even on the honeycomb lattice.
Thus, the resulting quantum spin-orbital liquid (QSOL)~\cite{Corboz2012}
cannot be a ``trivial'' featureless Mott insulator when the symmetry is not
broken spontaneously.

As explained in the above proof, the existence of a nontrivial generator
$I^0$ is important for this theorem.
In the case of $\alpha$-ZrCl$_3$ discussed in the main text, this element is not included
in the generators of the original $\mathrm{SU}(2) \times \mathrm{SU}(2)$ symmetry of the spin-orbital space,
but included in the emergent $\mathrm{SU}(4)$ symmetry in the strong spin-orbit coupling limit.
Thus, we can say that the $\mathrm{SU}(4)$ symmetry actually protects the nontrivial ground state of
the $\mathrm{SU}(4)$ Heisenberg model on the honeycomb lattice.

This proof of the LSMA theorem is not restricted to bosonic systems, and applies to both
bosonic and fermionic systems.  Thus, the generalization to the (zero-flux) $\mathrm{SU}(N)$-symmetric
Hubbard models is straightforward.
With $N$-flavor fermionic degrees of freedom in the $\mathrm{SU}(N)$ fundamental representation
at each site, the necessary condition for the existence of a featureless insulator is
that there exists a multiple of $N$ fundamental representations
per unit cell, which can form an $\mathrm{SU}(N)$ singlet.
We note that the LSMA theorem for $\mathrm{SU}(N)$ spin systems can be derived from the
$U\to \infty$ limit of the $\mathrm{SU}(N)$ Hubbard model at $1/N$ filling.
One can also extend the LSMA theorem to the systems with general representations
 on each site, starting from a Hubbard model.
That is, we include an appropriate onsite ``Hund'' coupling $J_H$ 
in the Hubbard model so that the desired representation have the lowest
energy, and then take the $J_H \to \infty$ limit afterwards.

The generalization to the 3D case with three translation operators,
$\trans_1,$ $\trans_2,$ and $\trans_3,$ is again straightforward and we will omit the
proof here, but it is useful to extend the
LSMA theorem to the case with a space group symmetry.
Recently, tighter constraints are obtained for nonsymmorphic space
group symmetries~\cite{PTAV2013,WPVZ2015} than what is implied by the
LSMA theorem based on the translation symmetries only.
This is because a nonsymmorphic symmetry behaves as a ``half'' translation, which would
reduce the size of the effective unit cell.

As a demonstration, here we only discuss the constraint given
by one nonsymmorphic (glide mirror or screw rotation) operation $\calG$,
by generalizing the flux insertion argument as in Ref.~\onlinecite{PTAV2013}.
We note that a tighter condition can be
derived by dividing the torus into the largest flat manifold, which is called
Bieberbach manifold, for some of the nonsymmorphic space groups~\cite{WPVZ2015}.

Among the 157 nonsymmorphic space groups, the 155 except for $I2_1 2_1
2_1$ (No. \textbf{24}) and $I2_1 3$ (No. \textbf{199}) include an
unremovable (essential) glide mirror or screw rotation symmetry
$\calG$~\cite{Konig1999}, so we will concentrate on these 155 to show
how $\calG$ works to impose a stronger constraint on filling.  The
nonsymmorphic operation $\calG$ consists of a point-group operation $G$
followed by a fractional (nonlattice) translation with a vector
$\bm{\alpha}$ in a direction left invariant by $G,$ \textit{i.e.} $\calG: \bm{r}
\mapsto G\bm{r}+\bm{\alpha}$ with $G \bm{\alpha}=\bm{\alpha}.$ We again
assume that the (possibly degenerate) ground states are separated from
the continuum of the excited states by a nonvanishing gap, and that the
gap does not collapse during the flux insertion process discussed below.
A ground state $\ket{\psi},$ which is $\mathrm{SU}(N)$-symmetric and has a
definite eigenvalue of all the crystalline symmetries including $\calG$
(\textit{i.e.} eigenstate of $\calG$), is chosen as the initial state.

We note that, for every nonsymmorphic space group except for $I2_1 2_1
2_1$ (No. \textbf{24}) and its key nonsymmorphic operation $\calG,$ we
can take an appropriate choice of primitive lattice vectors $\bm{a}_1,$
$\bm{a}_2,$ $\bm{a}_3$ with the following properties~\cite{WPVZ2015}:
(i) The associated translation $\bm{\alpha}$ is along the direction of
$\bm{a}_1$, and (ii) The plane spanned by $\bm{a}_2$ and $\bm{a}_3$ is
invariant under $G.$ Assuming this condition, we can show the tightest
condition derived from only one nonsymmorphic operation $\calG.$ For
simplicity, we consider the system consisting of $L_1 \times L_2 \times
L_3$ unit cells on a 3D torus (\textit{i.e.} impose the periodic boundary
conditions $\bm{r} \sim \bm{r} + L_\mu \bm{a}_\mu$
for $\mu=1,2,3$).

We take the smallest reciprocal lattice vector
$\tilde{\bm{k}}_1$ left invariant by $G,$
\textit{i.e.} $G\tilde{\bm{k}}_1=\tilde{\bm{k}}_1$ and
$\tilde{\bm{k}}_1$ generates the invariant sublattice of the
reciprocal lattice along $\tilde{\bm{k}}_1.$
We insert a flux on a torus by introducing a vector potential
$\bm{\calA}=\tilde{\bm{k}}_1/L_1.$
Since the ``magnetic flux'' reaches a multiple of $2\pi$
after this process because $\tilde{\bm{k}}_1$ is a reciprocal lattice vector,
the Hamiltonian of the system becomes equivalent to the initial one.
This happens precisely when the Hamiltonian is obtained from
the original Hamiltonian with a large gauge transformation.
The large gauge transformation to remove the inserted flux is
\begin{equation}
\calU_{\tilde{\bm{k}}_1} = \exp{\left[
\frac{i}{L_1} \sum_{\bm{r}} \tilde{\bm{k}}_1 \cdot \bm{r} I^0(\bm{r})
\right]}.
\end{equation}
Since $\bm{\calA}$ is left invariant under $\calG,$ the inserted flux
does not change the eigenvalues of $\calG.$  Thus, this phase factor due to the
large gauge transformation alone gives the
change of the eigenvalue of $\calG$ for $\ket{\psi}$ in the flux insertion process.
On the other hand,
\begin{equation}
	{\calG}^{-1} \calU_{\tilde{\bm{k}}_1} \calG
 \sim\calU_{\tilde{\bm{k}}_1}
e^{-(2\pi i \Phi_G (\tilde{\bm{k}}_1) n L_2 L_3 /N)},
\end{equation}
where $\Phi_G (\tilde{\bm{k}}_1)=\bm{\alpha} \cdot \tilde{\bm{k}}_1/(2\pi).$
For an unremovable glide or screw
symmetry, this phase factor has to be fractional.\footnote{We can show
that if $\Phi_G (\tilde{\bm{k}}_1)$ is an integer, then this nonsymmorphic
operation is removable, \textit{i.e.} can be reduced to a point-group operation
times a lattice translation by change of origin~\cite{Konig1999}.}
Thus, if we write $\Phi_G (\tilde{\bm{k}}_1)=p/\calS_G$ with $p,\calS_G$
relatively coprime, we can show a tighter bound for the filling
constraint to get a featureless Mott insulator without ground state
degeneracy because $\calS_G>1.$ In fact, to get a featureless Mott
insulator $pn L_2 L_3 /(N\calS_G)$ must at least be integer.  However,
if we choose $L_2$ and $L_3$ relatively prime to $N\calS_G,$ $n$ has to
be a multiple of $N\calS_G.$

If $n$ is not a multiple of $N\calS_G$ for some nonsymmorphic operation $\calG,$
this means the existence of degenerate ground states with a different eigenvalue
of $\calG,$ \textit{i.e.} implies the existence of gapless excitations or a gapped topological
order if the symmetry $\calG$ is not broken.  For example, in the case of the $\mathrm{SU}(4)$
Heisenberg model on the hyperhoneycomb lattice, $n=4,$ and the system can be trivial with
respect to the translation symmetry.  However, the space group of the hyperhoneycomb lattice
includes some nonsymmorphic operations, such as one glide mirror with $\calS_G=2.$
If we assume that nonsymmorphic symmetries are unbroken, the resulting QSOL (a possible
symmetric ground state) cannot be a trivial featureless Mott insulator.  Thus, we can say
this QSOL is protected by the nonsymmorphic space group symmetry of the lattice and
it can be called XSOL.

We note that as for the lattice (10,3)-$d,$ it is not enough to consider
only one symmetry operation and one has to consider the interplay of
multiple nonsymmorphic operations~\cite{PTAV2013}.
The derivation of
the tightest bound for all the 157 nonsymmorphic space groups with an
$\mathrm{SU}(N)$ symmetry is outside of the scope of this paper.
A nonsymmorphic symmetry sometimes exchanges
the bond label, and then it only exists when $J_\gamma$ obeys some
condition.  In this limited case, the generalized LSMA theorem only
applies in some parameter region defined by this condition.

\section{Examples of tricoordinated lattices}\label{trico}

The flux configurations for the 3D tricoordinated lattices listed in Table~\ref{lattice}
can be treated similarly to the Kitaev models on tricoordinated lattices~\cite{Kitaev2006,Obrien2016}
except for the difference in the gauge group.
Following Kitaev~\cite{Kitaev2006}, we use terminology of the lattice gauge theory.
The link variables $U_{ij}$ are Hermitian and unitary (in this case)
$4\times 4$ matrices defined for each bond (link) $\langle ij \rangle$ of the lattice. 
Each link variable depends on its type (color) of the bond as
\begin{equation}
        U_{ij} = \begin{cases}
    U^a = \tau^y \otimes I_2 & (\langle ij \rangle \in a) \\
    U^b = -\tau^x \otimes \sigma^z & (\langle ij \rangle \in b) \\
    U^c = -\tau^x \otimes \sigma^y & (\langle ij \rangle \in c)
  \end{cases}, \label{eq.lab}
\end{equation}
where $\bm{\tau}$ and $\bm{\sigma}$ are independent
Pauli matrices, following the original gauge (basis) used in Sec.~\ref{uij}
(not the one used in the previous section). The bond type $abc$ is determined from
which plane this bond belongs to in the same way as $\alpha$-ZrCl$_3.$
We note that in the 3D case we actually have six types of bonds with additional
$\pm 1$ factors, so $U_{ij} = \pm U^a,\,\pm U^b,\,\pm U^c$ depending on a detailed
structure of the bond $\langle ij \rangle.$ This comes from the spatial dependence of
the sign of the wavefunctions of the $d$-orbitals.
These additional $\pm 1$ factors can simply be gauged out as described in Ref.~\onlinecite{Yamada2018}.

In order to find a gauge transformation
to get an $\mathrm{SU}(4)$ Hubbard model, we have to check that every Wilson loop operator
is Abelian.  In an abuse of language, each Wilson loop will be called flux inside the loop.
We regard a Wilson loop operator $I_4$ as a zero flux, and $-I_4$ as a $\pi$ flux.
In order to get a desired gauge transformation,
it is enough to show that the flux inside every elementary loop $C$ is Abelian:
\begin{equation}
        \prod_{\langle ij \rangle \in C} U_{ij} = \zeta_C I_4, \label{Eq.fluxfree}
\end{equation}
with some phase factors $|\zeta_C|=1.$

Since $U_{ij}^2=I_4,$ not all the fluxes are independent. In the case of a $Z_2$ gauge
field, the constraints between multiple fluxes are called volume constraints~\cite{Obrien2016}.
However, due to the non-Abelian nature of the flux structure, it is subtle
whether they apply.  Fortunately, the above $U^\alpha$ ($\alpha=a,\,b,\,c$) obeys
the following anticommutation relations.
\begin{equation}
        \{U^\alpha,U^\beta\} = 2\delta^{\alpha\beta}I_4.
\end{equation}
This algebraic relation proves the product of the fluxes of the loops surrounding some
volume must vanish (volume constraints).  Moreover, we can easily show that, if every
bond color is used even times in each loop, which is a natural consequence for the lattices
admitting materials realization, the flux inside should always be Abelian with
$\zeta_C=\pm 1.$  Actually, every lattice included in Table~\ref{flux} obeys this
condition, so we have already proven all of them have an Abelian flux value.

The remaining subtle problem is which flux these elementary loops have, a zero flux,
or a $\pi$ flux. To check this, we need to investigate every loop one by one.
To calculate every flux value systematically, we often use space group symmetries to relate
two elementary loops, even though the system is in the strong spin-orbit
coupling limit. We note that the threefold rotation symmetry of the $xyz$-axes
of the Cartesian coordinate is not clear in the original gauge in Sec.~\ref{uij}.
This symmetry is important for some 3D models, although the spin quantization
axis along the (111) direction will make this symmetry explicit.
We have checked all the elementary loops in the tricoordinated lattices listed
here. In most cases, elementary loops of the same length
have the same flux due to some symmetry.
Only the flux value for the shortest elementary loops is shown in Table~\ref{flux}.

\begin{table}
        \centering
        \caption{\label{flux}Flux value of tricoordinated lattices.  Only the flux value for the shortest elementary loops is shown here.  Nonsymmorphic space group numbers are underlined.  NS means that nonsymmorphic symmetries of the lattice are enough to protect a quantum spin-orbital liquid state.  In addition to the contents of Table~\ref{lattice}, we also include O'Keeffe's three-letter codes~\cite{OKeeffe2003regular,OKeeffe2003semiregular}.}
        \begin{tabular}{ccccc}
                        Wells' & Lattice & O'Keeffe's & Minimal & Flux  \\
                        notation & name & code & loop length & value \\
                \hline
				(10,3)-$a$ & hyperoctagon & \textbf{srs} & 10 & 0-flux \\
                (10,3)-$b$ & hyperhoneycomb & \textbf{ths} & 10 & 0-flux \\
                (10,3)-$d$ & & \textbf{utp} & 10 & 0-flux \\
				nonuniform & $8^2.10$-$a$ & \textbf{lig} & 8 & $\pi$-flux \\
				(8,3)-$b$ & hyperhexagon & \textbf{etb} & 8 & $\pi$-flux \\
                nonuniform & stripyhoneycomb & \textbf{clh} & 6 & $\pi$-flux \\
				(6,3) & 2D honeycomb & \textbf{hcb} & 6 & $\pi$-flux
        \end{tabular}
\end{table}

\subsection{(10,3)-$a$}

\begin{figure}
\centering
\includegraphics[width=8.6cm]{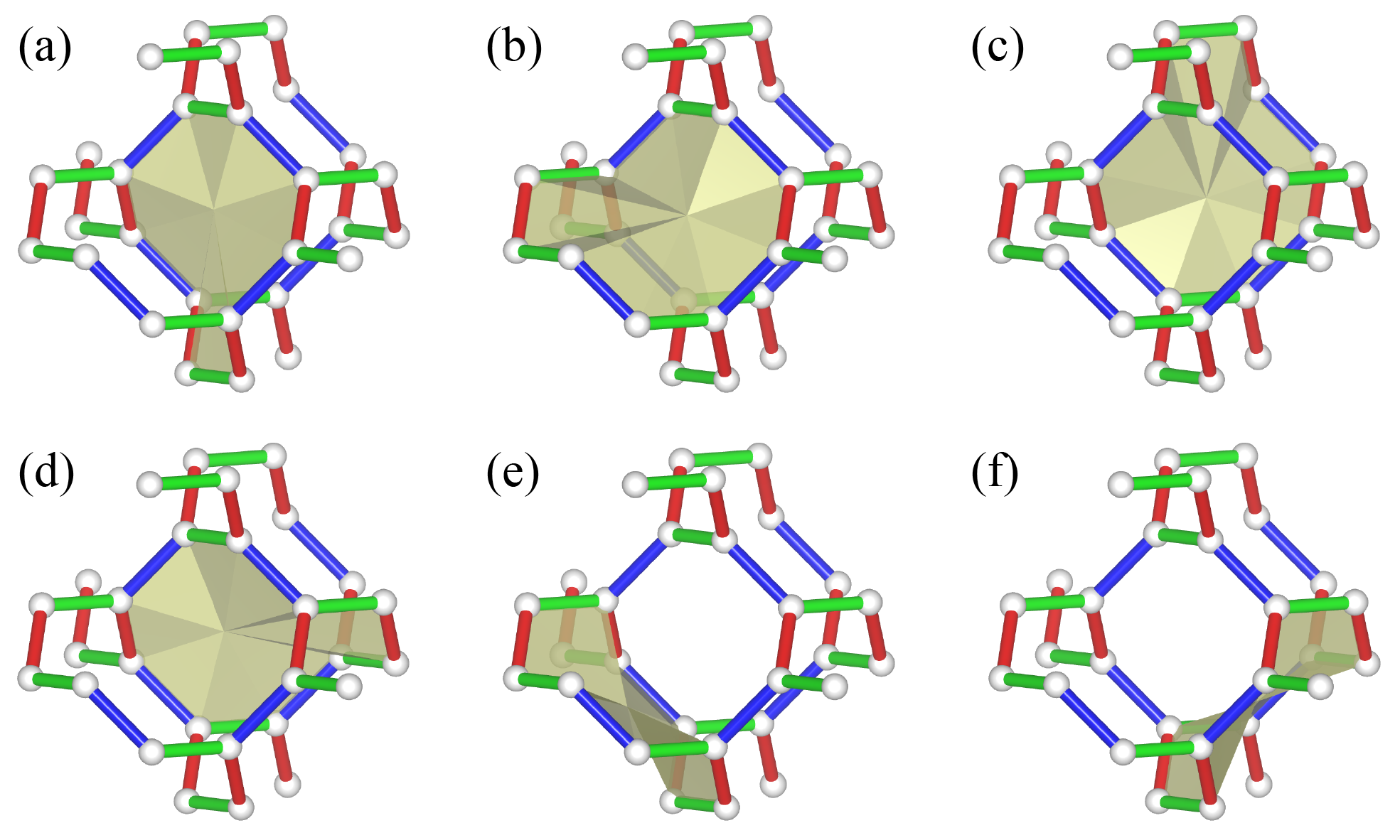}
\caption{Part of (10,3)-$a.$ All the six elementary loops~\cite{Hermanns2014} are highlighted by
yellow surfaces.  Loops (a)-(d) are related by the fourfold screw rotation,
and loops (e) and (f) are again related by the same symmetry.}
\label{10a}
\end{figure}

First of all, nonsymmorphic symmetries are useful to determine the flux value
because nonsymmorphic transformations often do not change the bond coloring
and effectively reduce the number of elementary loops.  As a concrete example,
we take the hyperoctagon lattice (10,3)-$a$ to show its usefulness.  (10,3)-$a$ has
six elementary loops of length 10~\cite{Hermanns2014}, and 4 of them are related by
the fourfold screw rotation symmetry [see Fig.~\ref{10a}(a)-(d)].
This fourfold screw exchanges the $b$-bonds for
the $c$-bonds, but this will not affect the flux value if the flux is Abelian
because the choice of the $xyz$-axes and its chirality is arbitrary.
The rest two elementary loops [see Fig.~\ref{10a}(e)-(f)] accidentally have the same coloring as
they are related by the screw symmetry.
Therefore, it is enough to check only two elementary loops, (a) and (e).
\begin{align}
U^cU^aU^cU^aU^bU^aU^cU^aU^cU^b &=I_4, \\
U^bU^aU^bU^aU^cU^aU^bU^aU^bU^c &=I_4.
\end{align}
From the above symmetry arguments, or from volume constraints,
we can conclude that all the six elementary loops
(of length 10) in (10,3)-$a$ have a zero flux.  This result agrees with the fact
that this zero-flux configuration is the unique $Z_2$ flux configuration
that obeys all the lattice symmetries of (10,3)-$a$~\cite{Obrien2016}.

\subsection{(10,3)-$b$}

\begin{figure}
\centering
\includegraphics[width=5cm]{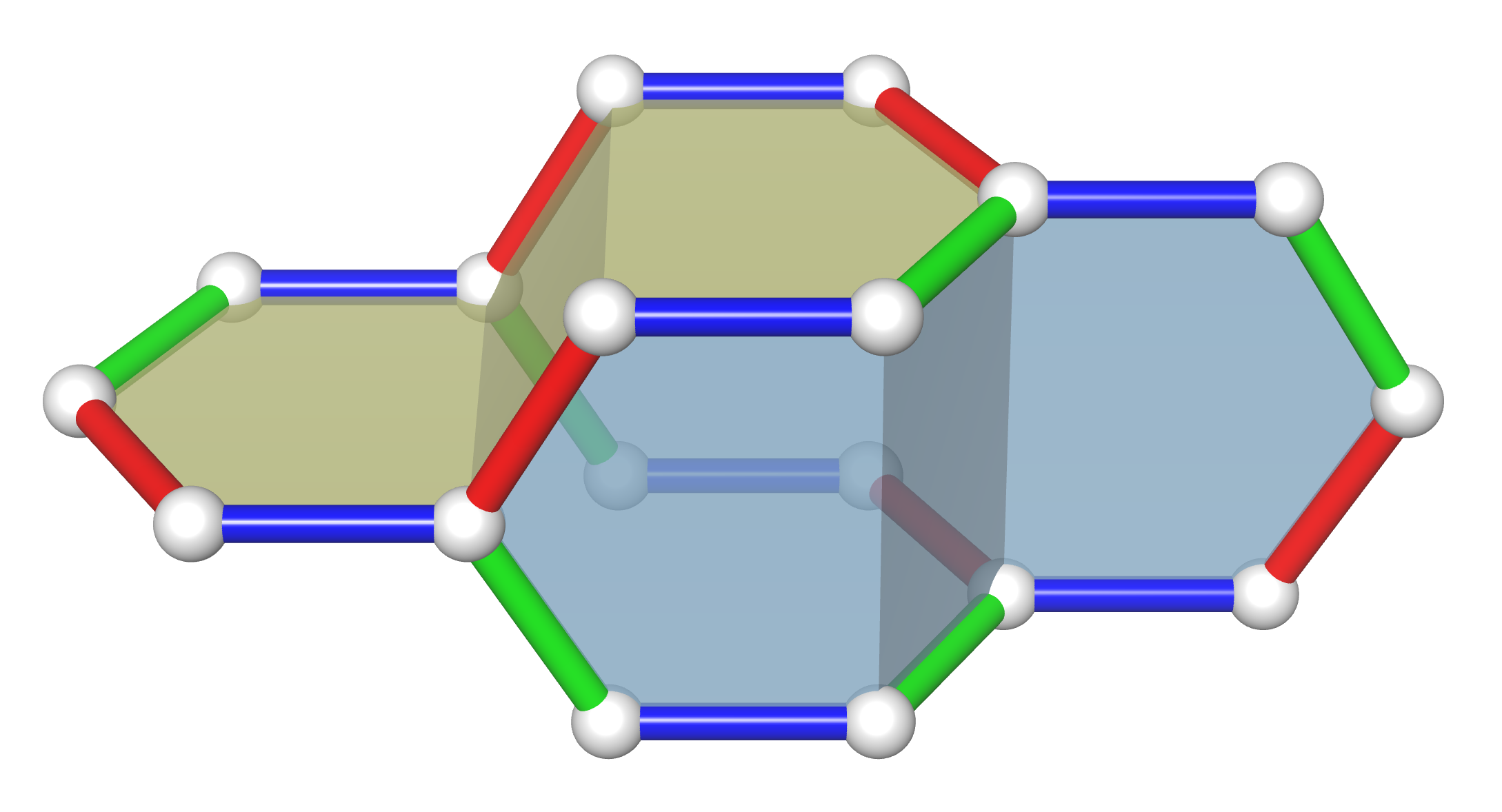}
\caption{Part of (10,3)-$b$ including four loops forming a volume constraint.
Two elementary loops with different coloring patterns are highlighted by
yellow and cyan surfaces, respectively.}
\label{10b}
\end{figure}

Among various point group symmetries, the inversion symmetry of the lattice
is the most useful.  As is the case in the honeycomb lattice, if an elementary
loop has an inversion center, then the flux inside this loop becomes the
square of some Pauli matrices times a complex number, which actually
only takes $1,i,-1,-i.$  Therefore, the existence of an inversion center
automatically proves that the flux is Abelian and should be $0$ or $\pi.$
This is another proof that a non-Abelian flux vanishes on some lattices.
This applies, for example, to the hyperhoneycomb lattice (10,3)-$b.$
All the four elementary loops of length 10 (10-loops) have an inversion center,
making the direct calculation easier.  We can classify these four 10-loops
into two pairs, where two loops are related by the glide mirror symmetry
with the same coloring pattern for each pair.  Therefore, it is enough to check
two loops, shown in the yellow and cyan surfaces, respectively, in Fig.~\ref{10b}.
\begin{align}
U^b U^c U^a U^c U^a U^b U^c U^a U^c U^a &= I_4. \\
U^a U^c U^b U^c U^b U^a U^c U^b U^c U^b &= I_4.
\end{align}
Therefore, all the four elementary loops in (10,3)-$b$ have a zero flux.

\subsection{(10,3)-$d$}

\begin{figure}
\centering
\includegraphics[width=6cm]{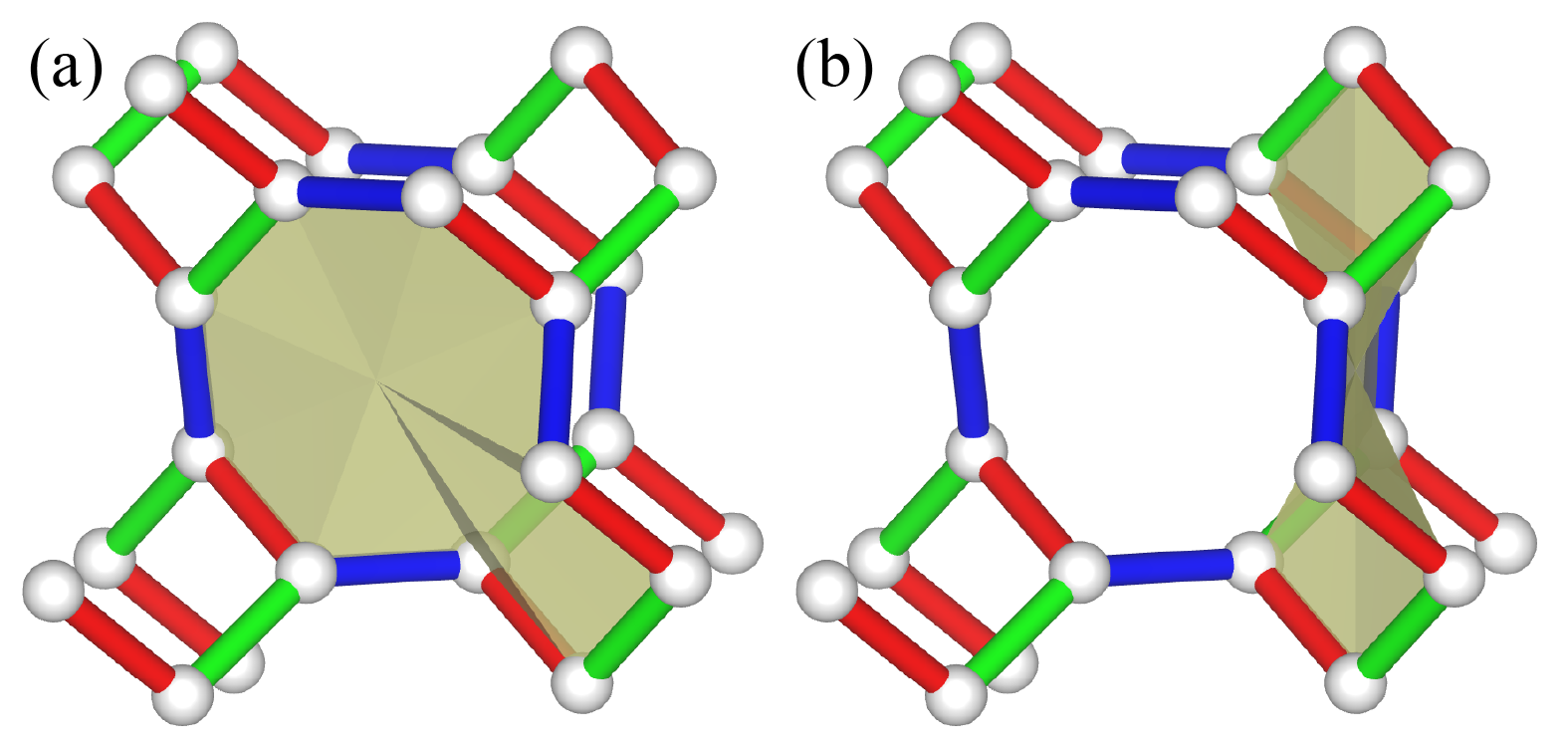}
\caption{Part of (10,3)-$d.$ (a) One of the type-A loops highlighted by the yellow surface.
(b) One of the type-B loops highlighted by the yellow surface.}
\label{10d}
\end{figure}

The structure of (10,3)-$d$ is related to (10,3)-$a$ because they share the same projection
onto the (001) plane, the 2D squareoctagon lattice.  Due to the difference in the chiralities
of the square spirals, the unit cell is enlarged in (10,3)-$d$ and possess 8 elementary loops
(of length 10) per unit cell.

Since this lattice does not allow any 120-degree configuration, we cannot simply
decide the bond coloring.  If we take the most symmetric bond coloring discussed in~\cite{Yamada2017XSL},
then the calculation becomes simple.  We can divide 8 elementary loops of length 10 into
two types.  Four type-A loops are spiraling up the octagon spiral
and then spiraling down the square spiral [see Fig.~\ref{10d}(a)].
All the four type-A loops are related by
the inversion symmetry or the twofold screw rotation symmetry
(the combination of them is the glide mirror symmetry), and thus have the same flux.
Four type-B loops are spiraling up the square spiral and then
spiraling down the nearest-neighbor square spiral [see Fig.~\ref{10d}(b)].
Four type-B loops are
related by the twofold screw rotation symmetry or by the glide mirror symmetry,
and have the same flux.  Thus, it is enough to check one for each type.
\begin{align}
U^b U^c U^a U^c U^a U^b U^a U^c U^a U^c &= I_4. \\
U^b U^a U^b U^a U^c U^b U^a U^b U^a U^c &= I_4.
\end{align}
The direct calculation tells us that the hopping model is in a zero-flux configuration.

\subsection{$8^2.10$-$a$}\label{810acolor}

\begin{figure}
\centering
\includegraphics[width=8.6cm]{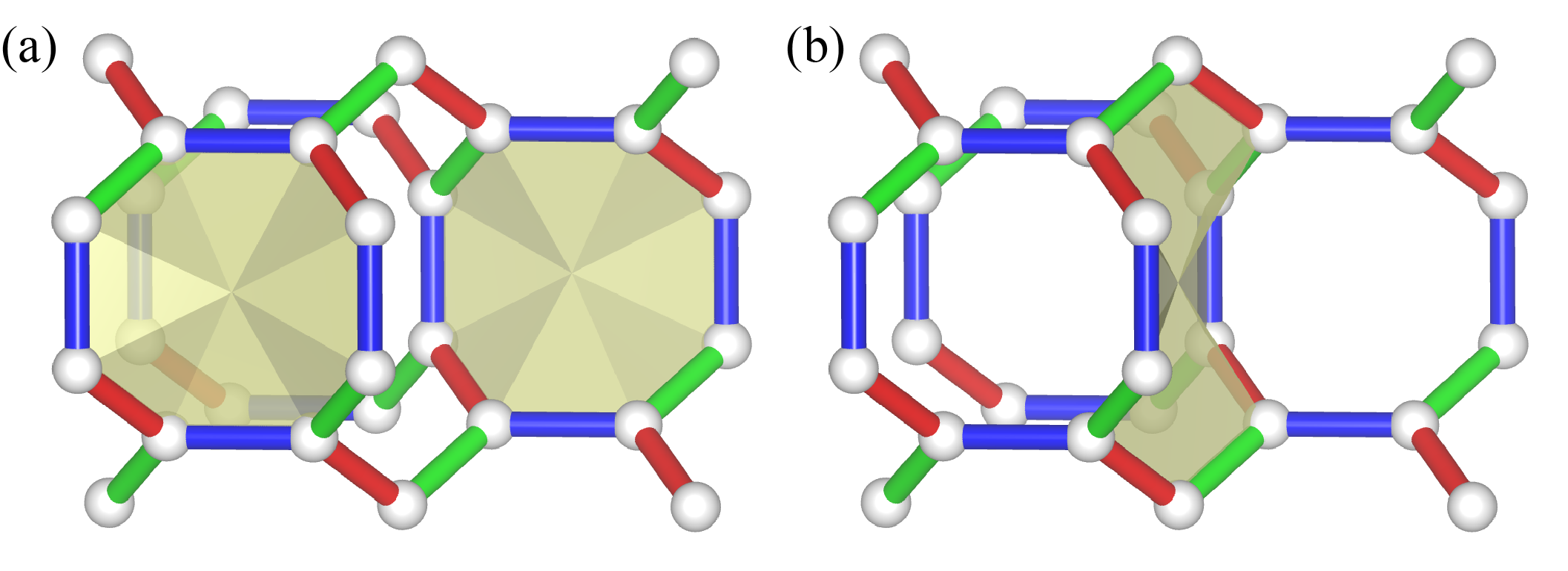}
\caption{Part of $8^2.10$-$a.$ (a) All the two 8-loops are shown by yellow
surfaces.  They are related by the fourfold screw rotation symmetry.
(b) One of the four 10-loops is shown by the yellow surface.
The rest are produced by applying the fourfold screw rotation around the
square spiral.}
\label{810a}
\end{figure}

$8^2.10$-$a$ is nonuniform, but Archimedean.  Therefore, each site is included in the
two types of elementary loops, some of length 8 and others of length 10.
The unit cell includes two elementary loops of length 8 (8-loops) [see Fig.~\ref{810a}(a)]
and four elementary loops of length 10 (10-loops) [see Fig.~\ref{810a}(b)].
It is enough to check one of the 8-loops and one of the 10-loops
because all the elementary loops of the same length are related by the fourfold screw
rotation symmetry.
\begin{align}
U^a U^c U^b U^c U^a U^c U^b U^c &= -I_4. \\
U^c U^a U^b U^a U^b U^c U^a U^b U^a U^b &= I_4.
\end{align}
Therefore, all the 8-loops have a $\pi$ flux and all the 10-loops have a zero flux.
We note that the hopping
model in this $\pi$-flux configuration does not break the original translation symmetry~\cite{Yamada2017XSL}.

\subsection{(8,3)-$b$}

\begin{figure}
\centering
\includegraphics[width=5cm]{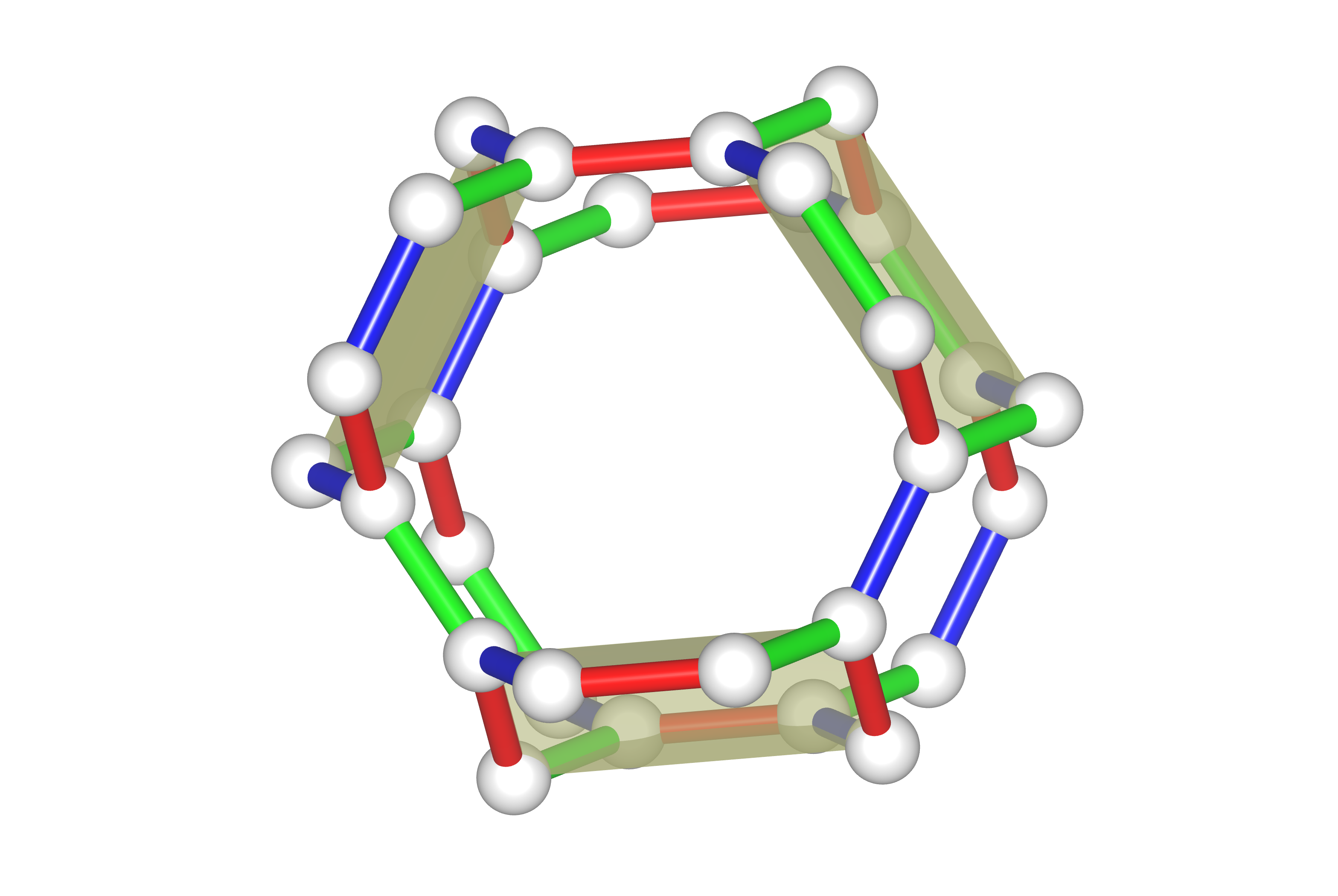}
\caption{Part of (8,3)-$b.$ All the three elementary loops of length 8 are
highlighted by yellow surfaces.  They are related by the threefold rotation symmetry.}
\label{8b}
\end{figure}

The hyperhexagon lattice (8,3)-$b$ has three elementary loops of length 8, and they
are related by the threefold rotation symmetry changing the $xyz$-axes, as shown in
Fig.~\ref{8b}.  Therefore, it is
enough to check only one of them.  The direct calculation tells us that it has a $\pi$ flux.
\begin{equation}
U^a U^c U^b U^c U^a U^c U^b U^c = -I_4.
\end{equation}

Therefore, (8,3)-$b$ is in the $\pi$-flux configuration. We note that there is another
elementary loop of length 12, but the flux value is immediately determined to
be zero due to the accidental fourfold symmetry of the coloring.
It is worth mentioning the hopping
model in this $\pi$-flux configuration does not break the original translation symmetry,
and thus the LSMA theorem applies as it is to the $\pi$-flux $\mathrm{SU}(4)$ Hubbard model, as well
as the $\mathrm{SU}(4)$ Heisenberg model.

\subsection{Stripyhoneycomb lattice}

\begin{figure}[b]
\centering
\includegraphics[width=8.6cm]{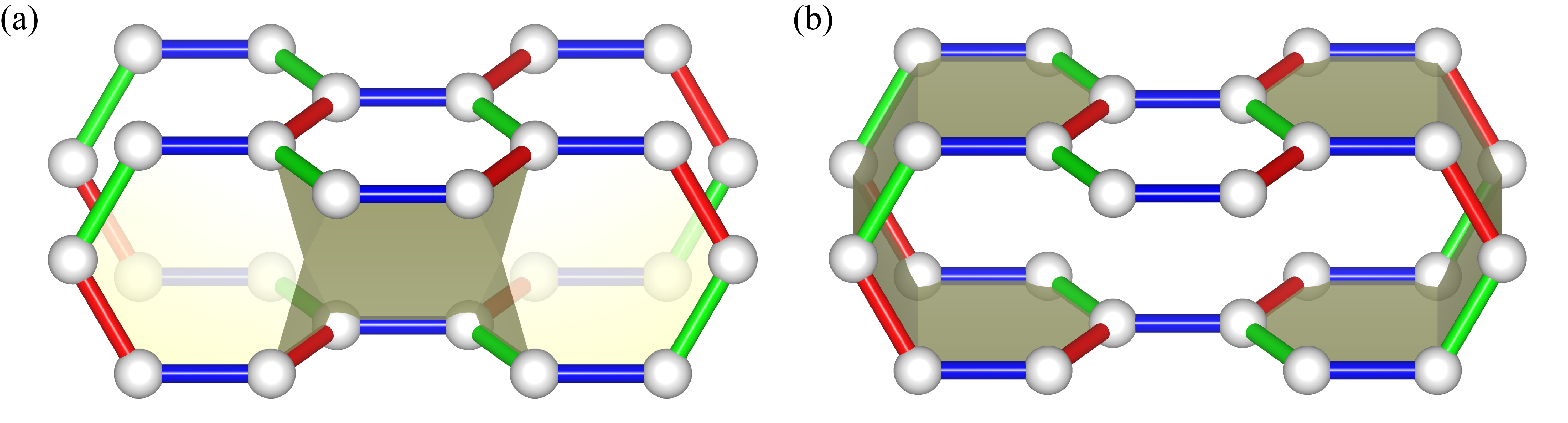}
\caption{Part of the stripyhoneycomb lattice. (a) A loop of length 14 is highlighted.
(b) A pair of loops of length 12 are highlighted. They are related by the inversion
symmetry (or the volume constraint) and thus have the same flux.}
\label{stripy}
\end{figure}

The stripyhoneycomb lattice is nonuniform, so the length of the shortest elementary loops
differs in space.  Every elementary loop of length 6 is the same as the honeycomb, and thus
has a $\pi$ flux.  The structure includes two types of the $\pi$-flux hexagons aligning
in different planes~\cite{Kimchi2014}.
In addition, there exist a long loop of length 14 (14-loop) and a twisted
loop of length 12 (12-loop) [see Fig.~\ref{stripy}].
These four types of elementary loops are enough to determine the flux values.

One 14-loop shown in Fig.~\ref{stripy}(a) has a zero flux because
\begin{equation}
		U^a U^c U^a U^b U^c U^b U^c U^a U^c U^a U^b U^c U^b U^c = I_4.
\end{equation}

One 12-loop shown on the right-hand side of Fig.~\ref{stripy}(b) also has a zero flux because
\begin{equation}
U^a U^b U^c U^a U^b U^c U^b U^a U^c U^b U^a U^c = I_4.
\end{equation}

There are many other tricoordinated lattices not discussed in this paper,
so it is future work to determine the flux values for all the possible tricoordinated
lattices.

\bibliography{paper}

\end{document}